\documentclass[aps,prb,twocolumn,superscriptaddress,floatfix,showpacs,10pt,letterpaper]{revtex4}
\usepackage{graphicx}
\usepackage{dcolumn}
\usepackage{bm}
\usepackage{subfigure}
\usepackage[USenglish]{babel}
\usepackage{amssymb,amsmath}
\usepackage{relsize}
\usepackage{lmodern}
\usepackage{scalefnt}
\usepackage{tikz}
\usetikzlibrary{arrows,calc,scopes}
\usetikzlibrary{shapes.geometric}
\allowdisplaybreaks[1]

\usepackage[colorlinks,citecolor=blue]{hyperref}

\newcommand{\be}{\begin{equation}} \newcommand{\ee}{\end{equation}}
\newcommand{\bse}{\begin{subequations}}\newcommand{\ese}{\end{subequations}}
\newcommand{\bpm}{\begin{pmatrix}} \newcommand{\epm}{\end{pmatrix}}
\newcommand{\bmm}{\begin{matrix}} \newcommand{\emm}{\end{matrix}}

\newcommand{\Z}{\mathbb{Z}}

\newcommand{\R}{\mathbb{R}}

\renewcommand{\v}[1]{\boldsymbol{#1}} \renewcommand{\t}[1]{\tilde{#1}}

\newcommand{\e}{\hspace{1pt}\mathrm{e}}
\newcommand{\dd}{\hspace{1pt}\mathrm{d}}
\newcommand{\imth}{\hspace{1pt}\mathrm{i}\hspace{1pt}}

\newcommand{\Ref}[1]{Ref.~\onlinecite{#1}}

\newcommand{\eq}[1]{(\ref{#1})} \newcommand{\eqn}[1]{eqn.~(\ref{#1})}

\newcommand{\<}{\langle} \renewcommand{\>}{\rangle}
 
\newcommand{\Tr}{{\rm Tr}}    
 \newcommand{\prt}{\partial}

\newcommand{\ie}{{\it ie~}}  \newcommand{\etc}{{\it
etc~}}

\newcommand{\al}{\alpha}  
 
 \newcommand{\ga}{\gamma}
  \newcommand{\la}{\lambda}
 \newcommand{\om}{\omega} 
\renewcommand{\th}{\theta}  \newcommand{\si}{\sigma}

 \newcommand{\cH}{ {\cal H} }  \newcommand{\cL}{ {\cal L} } 
   
\newcommand{\cZ}{ {\cal Z} }

\begin{document}


\begin{titlepage}

\title{Quantized topological terms in weak-coupling gauge theories with
a global symmetry\\
and their connection to symmetry enriched topological phases}

\author{Ling-Yan Hung}
\affiliation{Department of Physics, Harvard University, Cambridge MA 02138}
\affiliation{Perimeter Institute for Theoretical Physics, 31 Caroline St N,
Waterloo, ON N2L 2Y5, Canada}
\author{Xiao-Gang Wen}
\affiliation{Perimeter Institute for Theoretical Physics, 31 Caroline St N,
Waterloo, ON N2L 2Y5, Canada}
\affiliation{Department of Physics, Massachusetts Institute of Technology,
Cambridge, Massachusetts 02139, USA}
\affiliation{Institute for Advanced Study, Tsinghua University, Beijing,
100084, P. R. China}

\date{\today}

\begin{abstract}
We study the quantized topological terms in a weak-coupling gauge theory with
gauge group $G_g$ and a global symmetry $G_s$ in $d$ space-time dimensions. We
show that the  quantized topological terms are classified by a pair
$(G,\nu_d)$, where $G$ is an extension of $G_s$ by $G_g$ and $\nu_d$ an element
in group cohomology $\cH^d(G,\R/\Z)$.  When $d=3$ and/or when $G_g$ is finite,
the  weak-coupling gauge theories with  quantized topological terms describe
gapped symmetry enriched topological (SET) phases (\ie gapped long-range
entangled phases with symmetry).  Thus, those SET phases are classified by
$\cH^d(G,\R/\Z)$, where $G/G_g=G_s$.  We also apply our theory to a simple case
$G_s=G_g=Z_2$, which leads to 12 different SET phases in 2+1D, where quasiparticles have
different patterns of fractional $G_s=Z_2$ quantum numbers and fractional
statistics.  If the  weak-coupling gauge theories are gapless, then the
different quantized topological terms may describe different gapless phases of
the gauge theories with a symmetry $G_s$, which may lead to different
fractionalizations of $G_s$ quantum numbers and different fractional statistics
(if in 2+1D).

\end{abstract}

\maketitle

\end{titlepage}

{\small \setcounter{tocdepth}{1} \tableofcontents }

\section{Introduction}

For a long time, we thought that Landau symmetry breaking
theory\cite{L3726,GL5064,LanL58} describes all phases and phase transitions.
In 1989, through a theoretical study of high $T_c$ superconducting model, we
realized that there exists a new kind of orders -- topological order -- which
cannot be described by Landau symmetry breaking theory.\cite{Wtop,WNtop,Wrig}
Recently, it was found that topological orders are related to long range
entanglements.\cite{LW0605,KP0604} In fact, we can regard topological order as
pattern of long range entanglements\cite{CGW1038} defined through local unitary
(LU) transformations.\cite{LW0510,VCL0501,V0705}

\begin{figure}[b]
\begin{center}
\includegraphics[scale=0.49]{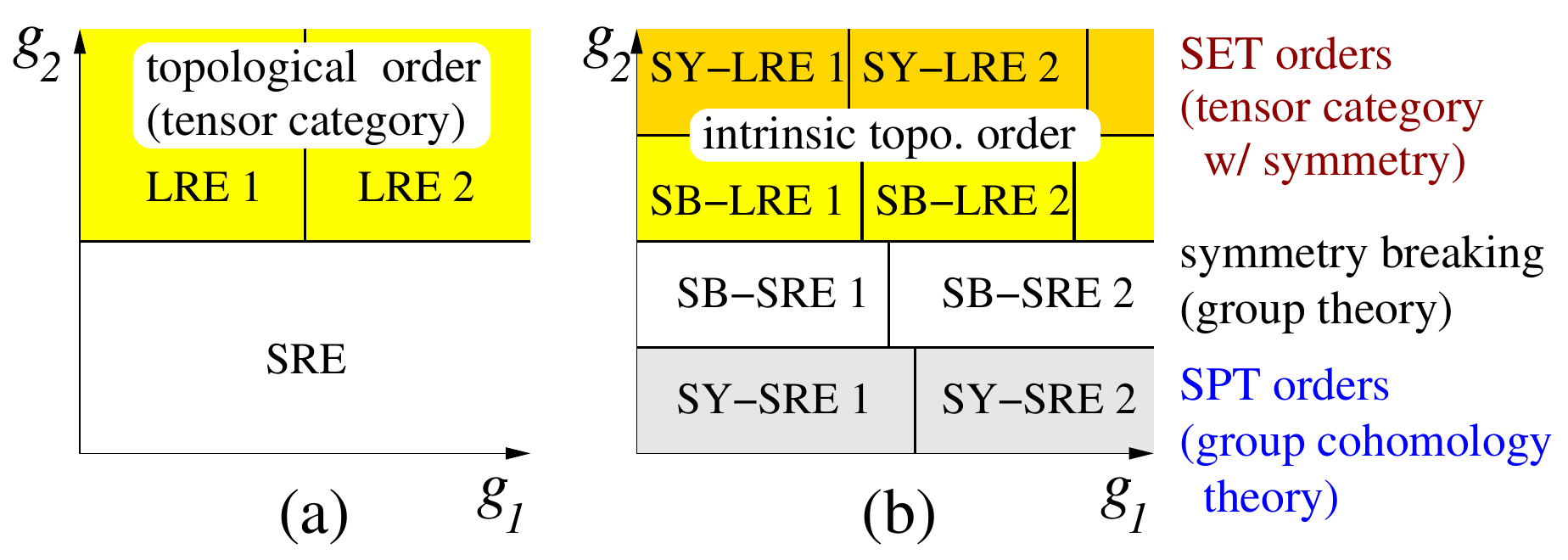}
\end{center}
\caption{
(Color online)
(a) The possible gapped phases for a class of Hamiltonians $H(g_1,g_2)$ without
any symmetry.  (b) The possible gapped phases for the class of
Hamiltonians $H_\text{symm}(g_1,g_2)$ with a symmetry.  The yellow regions in
(a) and (b) represent the phases with long range entanglement.  Each phase is
labeled by its entanglement properties and symmetry breaking properties.  SRE
stands for short range entanglement, LRE for long range entanglement, SB for
symmetry breaking, SY for no symmetry breaking.  SB-SRE phases are the Landau
symmetry breaking phases.  The SY-SRE phases are the SPT phases.  The SY-LRE
phases are the SET phases.
}
\label{topsymm}
\end{figure}

The notion of topological orders and long range entanglements leads to a more
general and also more detailed picture of phases and phase transitions (see
Fig.  \ref{topsymm}).\cite{CGW1038} For gapped quantum systems without any
symmetry, their quantum phases can be divided into two classes: short range
entangled (SRE) states and long range entangled (LRE) states.

SRE states are states that can be transformed into direct product states via LU
transformations. All SRE states can be transformed into each other via  LU
transformations. So all SRE states belong to the same phase (see Fig.
\ref{topsymm}a), \ie all SRE states can continuously deform into each other
without closing energy gap and without phase transition.

LRE states are states that cannot be transformed into  direct product states
via LU transformations.  It turns out that, in general, different LRE states
cannot be connected to each other through LU transformations. The LRE states
that are not connected via LU transformations
represent different quantum phases.  Those different quantum phases are nothing
but the topologically ordered phases.

Chiral spin liquids,\cite{KL8795,WWZ8913}
fractional quantum Hall states\cite{TSG8259,L8395},
$Z_2$ spin liquids,\cite{RS9173,W9164,MS0181}
non-Abelian fractional quantum Hall states,\cite{MR9162,W9102,WES8776,RMM0899}
\etc are examples of topologically ordered phases.  The mathematical foundation
of topological orders is closely related to tensor category
theory\cite{FNS0428,LW0510,CGW1038,GWW1017} and simple current
algebra.\cite{MR9162,LWW1024} Using this point of view, we have developed a
systematic and quantitative theory for non-chiral topological orders in 2D
interacting boson and fermion systems.\cite{LW0510,CGW1038,GWW1017} Also for
chiral 2D topological orders with only Abelian statistics, we find that we can
use integer $K$-matrices to describe them.\cite{BW9045,R9002,FK9169,WZ9290,BM0535,KS1193}

For gapped quantum systems with symmetry, the structure of phase diagram is
even richer (see Fig. \ref{topsymm}b).  Even SRE states now can belong to
different phases.  One class of non-trivial SRE phases for Hamiltonians with
symmetry  is the Landau symmetry breaking states.  But even SRE states that do
not break the symmetry of the Hamiltonians can belong to different phases.  The
1D Haldane phase for spin-1 chain\cite{H8364,AKL8877,GW0931,PBT0959} and
topological insulators\cite{KM0501,BZ0602,KM0502,MB0706,FKM0703,QHZ0824} are
non-trivial examples of phases with short range entanglements that do not break
any symmetry.  We will call this kind of phases SPT phases.  The term ``SPT
phase'' may stand for Symmetry Protected Topological
phase,\cite{GW0931,PBT0959} since the known examples of those phases, the
Haldane phase and the topological insulators, were already referred as
topological phases. The term ``SPT phase'' may also stand for Symmetry
Protected Trivial phase, since those phases have no long range entanglements
and have trivial topological orders.

It turns out that there is no gapped bosonic LRE state in 1D.\cite{VCL0501} So
all 1D gapped bosonic states are either symmetry breaking states or SPT states.
This realization led to a complete classification of all 1D gapped bosonic
quantum phases.\cite{CGW1107,SPC1139,CGW1128}

In \Ref{CLW1141,CGL1172}, the classification of 1D SPT phase is generalized to
any dimensions: \emph{For gapped bosonic systems in $d$ space-time dimensions
with an on-site symmetry $G_s$, we can construct distinct SPT phases that do
not break the symmetry $G_s$ from the distinct elements in $\cH^{d}[G_s, U(1)]$
-- the $d$-cohomology class of the symmetry group $G_s$ with $U(1)$ as
coefficient.} We see that we have a quite systematic understanding of SRE
states with symmetry.\cite{LS0903,LV1219}

For gapped LRE states with symmetry, the possible quantum phases should be much
richer than SRE states. We may call those phases Symmetry Enriched Topological
(SET) phases.  Projective symmetry group (PSG) was introduced to study the SET
phases.\cite{W0213,W0303a,WV0623} The PSG describes how the quantum numbers of
the symmetry group $G_s$ get fractionalized on the gauge
excitations.\cite{W0303a} When the gauge group $G_g$ is Abelian, the PSG
description of the SET phases can be be expressed in terms of group cohomology:
The different SET states with symmetry $G_s$ and gauge group $G_g$ can be
(partially) described by $\cH^2(G_s,G_g)$.\cite{EH1293} Many examples of the
SET states can be found in \Ref{W0213,KLW0834,KW0906,LS0903,YFQ1070}.

Recently,  Mesaros and Ran proposed a quite systematic understanding of a
subclass of SET phases:\cite{MR1235} One can use the elements of
$\cH^d(G_s\times G_g,\R/\Z)$ to describe the SET phases in $d$ space-time
dimensions with a finite gauge group $G_g$ and a finite global symmetry group
$G_s$.  Here $\cH^d(G_s\times G_g,\R/\Z)$ is the group cohomology class of
group $G_s\times G_g$.  This result is based on the group cohomology theory of
the SPT phases\cite{CGL1172} and the Levin-Gu duality between the SPT phases
and the ``twisted'' weak-coupling gauge theories.\cite{LG1220,S1276,HW1267}
Also, Essin and Hermele generalized the results of
\Ref{W0213,W0303a,KLW0834,KW0906} and studied quite systematically the SET
phases described by a $G_g=Z_2$ gauge theory.\cite{EH1293}  They show that some
of those SET phases can be classified by $\cH^2(G_s,G_g)$.

In this paper, we will develop a somewhat systematic understanding of SET
phases, following a path-integral approach developed for the  group cohomology
theory of the SPT phases\cite{CGL1172} and the topological gauge
theory.\cite{DW9093,HW1267} The idea is to classify quantized topological terms
in weak-coupling gauge theory with symmetry.  If the  weak-coupling gauge
theory happens to have a gap, then the different quantized topological terms
will describe different SET phases.  This allows us to obtain and generalize
the results in \Ref{MR1235,EH1293}.  Since weak-coupling gauge theories only
describe some topological ordered states, our theory only describes some of the
SET states.

We show that  quantized topological terms  in symmetric  weak-coupling gauge
theory in $d$ space-time dimensions with a gauge group $G_g$ and a global
symmetry group $G_s$ can be described by a pair $(G,\nu_d)$, where $G$ is an
extension of $G_s$ by $G_g$ and $\nu_d$ is an element in $\cH^d(G,\R/\Z)$.  (An
extension of $G_s$ by $G_g$ is group $G$ that contain $G_g$ as a normal
subgroup and satisfy $G/G_g=G_s$.) When $G_g$ is finite or when $d=3$, the
weak-coupling gauge theory is gapped. In this case, $(G,\nu_d)$ describe
different SET phases.  Note that the extension $G$ is nothing but the PSG
introduced in \Ref{W0213}.  Also, when the symmetry group $G_s$ contains
anti-unitary transformations, those anti-unitary transformations will act
non-trivially on $\R/\Z$: $x \to -x$, $x\in \R/\Z$.\cite{CGL1172}

In appendix \ref{lab}, we will show that we can use $(y_0,y_1,...,y_d)$ with
\begin{align}
 y_k\in \cH^k[G_s,\cH^{d-k}(G_g,\R/\Z)]
\end{align}
to  label the  elements in $\cH^d(G,\R/\Z)$.  However, such a labeling may not
be one-to-one and it may happen that only some of $(y_0,y_1,...,y_d)$
correspond to  the  elements in $\cH^d(G,\R/\Z)$.  But for every element in
$\cH^d(G,\R/\Z)$, we can find a $(y_0,y_1,...,y_d)$ that corresponds to it.  If
we choose a special extension $G=G_g\times G_s$, then we recover the result in
\Ref{MR1235} if $G$ is finite: a set of SET states can be can be described by
$(y_0,y_1,...,y_d)$ with an one-to-one correspondence (see \eqn{GsGg}):
\begin{align}
\cH^{d}(G_s\times G_g,\R/\Z)
& =
 \oplus_{p=0}^{d} \cH^{d-p}[G_s,\cH^{p}(G_g,\R/\Z)]
\nonumber\\ &=
 \oplus_{p=0}^{d} \cH^{d-p}[G_g,\cH^{p}(G_s,\R/\Z)]
.
\end{align}
The term $\cH^{d}[G_s,\cH^{0}(G_g,\R/\Z)]=\cH^{d}(G_s,\R/\Z)$ describes the
quantized topological terms associated with only the symmetry $G_s$ which
describe the SPT phases.  The term
$\cH^{0}[G_s,\cH^{d}(G_g,\R/\Z)]=\cH^{d}(G_g,\R/\Z)$ describes the quantized
topological terms associated with pure gauge theory.  Other terms
$\oplus_{p=1}^{d-1} \cH^{d-p}[G_s,\cH^{p}(G_g,\R/\Z)]$ describe the quantized
topological terms that involve both gauge theory $G_g$ and symmetry $G_s$.
Those terms describe how $G_s$ quantum numbers get fractionalized on gauge-flux
excitations.\cite{MR1235}

When $G_g$ is Abelian, the different extensions, $G$,  of $G_s$ by $G_g$ is
classified by $\cH^2(G_s,G_g)$.  This reproduces a result in \Ref{EH1293}.

\section{A simple formal approach}

First let us describe a simple formal approach that allows us to quickly obtain
the above results.  We know that the SPT phases in $d$-dimensional discrete
space-time are described by topological non-linear $\si$-models with symmetry
$G$:
\begin{align}
 \cL= \frac{1}{\la_s} [\prt g(x^\mu)]^2 + \imth W_\text{top}(g), \ \ g\in G
\end{align}
where $\la_s\to \infty$, and the $2\pi$-quantized topological term $\int
W_\text{top}(g)$ is given by an element in $\cH^{d}(G,\R/\Z)$. Different
elements in $\cH^{d}(G,\R/\Z)$ describe different SPT phases.\cite{CGL1172}  If
we ``gauge'' the symmetry $G$, the topological non-linear $\si$-model will
become a gauge theory:
\begin{align}
 \cL= \frac{1}{\la_s} [(\prt -\imth A) g(x^\mu)]^2 + \imth W_\text{top}(g,A)
+\frac{(F_{\mu\nu})^2}{\la},
\end{align}
where $W_\text{top}(g,A)$ is the gauged topological term.  For those
topological term that can be expressed in continuous field theory,
$W_\text{top}(g,A)$ can be obtained from $W_\text{top}(g)$ by replacing
$\prt_\mu$ by $\prt_\mu-\imth A_\mu$.  When $G_s$ and $G_g$ are finite,
$W_\text{top}(g,A)$ can be constructed explicitly in discrete
space-time.\cite{ZW}

If we further integrate out $g$, we will get a pure gauge theory with a
topological term
\begin{align}
 \cL= \frac{(F_{\mu\nu})^2}{\la} +\imth W_\text{top}(A)
.
\end{align}
This line of thinking suggests that the quantized topological term $\int \t
W_\text{top}(A)$ in symmetric gauge theory is classified by the same
$\cH^{d}(G,\R/\Z)$ that classifies the $2\pi$-quantized topological term
$\int W_\text{top}(g)$.

Now let us consider topological non-linear $\si$-models with symmetry
$G_s\times G_g$:
\begin{align}
 \cL= \frac{1}{\la_s} [\prt g(x^\mu)]^2 + \imth W_\text{top}(g), \ \
g\in G=G_s\times G_g,
\end{align}
where the $2\pi$-quantized topological term $\int W_\text{top}(g)$ is
classified by $\cH^{d}(G_s\times G_g,\R/\Z)$.  If we ``gauge'' only a
subgroup $G_g$ of the total symmetry group $G_s\times G_g$, we will get a gauge
theory:
\begin{align}
\cL= \frac{1}{\la_s} [(\prt -\imth A) g(x^\mu)]^2 + \imth
W_\text{top}(g,A) +\frac{(F_{\mu\nu})^2}{\la}
\end{align}
with global symmetry $G_s$. This line of thinking suggests that the
quantized topological term $\int W_\text{top}(g,A)$ is classified by the
same $\cH^{d}(G_s\times G_g,\R/\Z)$.

We can generalize the above approach to obtain more general quantized
topological terms in weak-coupling gauge theory with gauge group $G_g$ and
symmetry $G_s$.  We start with a group $G$ which is an extension of the
symmetry group $G_s$ by the gauge group $G_g$:
\begin{align}
 1\to G_g \to G \to G_s \to 1.
\end{align}
In other words, $G$ contains a normal subgroup $G_g$ such that $G/G_g=G_s$.  So
we can start with a topological non-linear $\si$-models with symmetry $G$:
\begin{align}
\label{nltGG}
 \cL= \frac{1}{\la_s} [\prt g(x^\mu)]^2 + \imth W_\text{top}(g), \ \ g\in G,
\end{align}
where the $2\pi$-quantized topological term $\int W_\text{top}(g)$ is
classified by $\cH^{d}(G,\R/\Z)$.  If we ``gauge'' only a subgroup $G_g$ of
the total symmetry group $G$, we will get a gauge theory:
\begin{align}
\label{nltGGg}
 \cL= \frac{1}{\la_s} [(\prt -\imth A) g(x^\mu)]^2 + \imth W_\text{top}(g,A)
+\frac{(F_{\mu\nu})^2}{\la}
\end{align}
with global symmetry $G_s=G/G_g$. This line of thinking suggests that the
quantized topological term $\int W_\text{top}(g,A)$ is classified by
$\cH^{d}(G,\R/\Z)$.

So more generally, the SET states in $d$-dimensional space-time with gauge
group $G_g$ and symmetry group $G_s$ are labeled by the elements in
$\cH^{d}(G,\R/\Z)$, where $G$ the extension of the symmetry group $G_s$ by the
gauge group $G_g$, provided that the symmetric gauge theory \eq{nltGG} is gapped
in small $\la$ limit and $d \geq 3$.  If the symmetric gauge \eqn{nltGG} is
gapless in small $\la$ limit, then $\cH^{d}(G,\R/\Z)$ describes different
gapless phases of the symmetric gauge theory.

The above approach is formal and hand-waving.  When $G$ is finite, we can
rigorously obtain the above results, which is described in \Ref{ZW}.  In the
following, we will discuss such an approach assuming $G_g$ is finite (but $G_s$
can be finite or continuous).  Then we will discuss another approach that allows
us to obtain the above result more rigorously for the case $G=G_s\times G_g$
where $G_s,G_g$ can be finite or continuous.

\section{An exact approach for finite $G_g$}

This approach is based on the formal approach \eq{nltGGg} discussed above, where
$G$ is an extension of the symmetry group $G_s$ by the gauge group $G_g$:
$G/G_g=G_s$. We will make the above approach exact by putting the theory on
space-time lattice of $d$ dimensions.  


\subsubsection{Discretize space-time}
\label{disltgauge}

We will discretize the space-time $M$ by considering its
triangulation $M_\text{tri}$ and define the $d$-dimensional
gauge theory on such a triangulation.  We will call such a theory a lattice
gauge theory.  We will call the triangulation $M_\text{tri}$ a
space-time complex, and a cell in the complex a simplex.

\begin{figure}[tb] 
\begin{center} 
\includegraphics[scale=0.6]{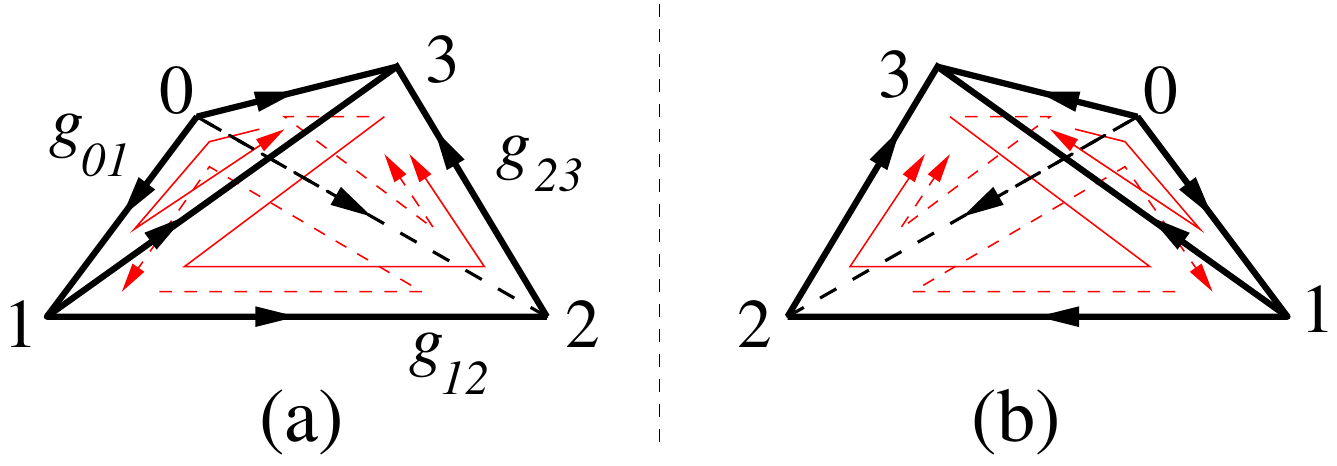} \end{center}
\caption{ (Color online) Two branched simplices with opposite orientations.
(a) A branched simplex with positive orientation and (b) a branched simplex
with negative orientation.  } 
\label{mir} 
\end{figure}

In order to define a generic lattice theory on the space-time complex
$M_\text{tri}$, it is important to give the  vertices of each simplex a local
order.  A nice local scheme to order  the  vertices is given by a branching
structure.\cite{C0527,CGL1172} A branching structure is a choice of orientation
of each edge in the $d$-dimensional complex so that there is no oriented loop
on any triangle (see Fig. \ref{mir}).

The branching structure induces a \emph{local order} of the vertices on each
simplex.  The first vertex of a simplex is the vertex with no incoming edges,
and the second vertex is the vertex with only one incoming edge, \etc.  So the
simplex in  Fig. \ref{mir}a has the following vertex ordering: $0,1,2,3$.

The branching structure also gives the simplex (and its sub simplexes) an
orientation denoted by $s_{ij...k}=1,*$.  Fig. \ref{mir} illustrates two
$3$-simplices with opposite orientations $s_{0123}=1$ and $s_{0123}=*$.  The
red arrows indicate the orientations of the $2$-simplices which are the
subsimplices of the $3$-simplices.  The black arrows on the edges indicate the
orientations of the $1$-simplices.

\subsubsection{Gauged non-linear $\si$-model on space-time lattice}

To put \eq{nltGGg} on space-time lattice, we put the $g(x^\mu)\in G$ field on
the vertices of the space-time complex, which becomes $g_i$ where $i$ labels
the vertices.  We also put the gauge field on the edges $ij$ which becomes
$g_{ij}\in G_g$.  

The action amplitude for a $d$-cell $(ij...k)$ is complex function of
$g_i$ and $g_{ij}$: $V_{ij...k}(\{g_{ij}\},\{g_i\})$
The partition function is given by
\begin{align}
 Z=\sum_{ \{g_{ij}\},\{g_i\} }
\prod_{(ij...k)} [V_{ij...k}(\{g_{ij}\},\{g_i\})]^{s_{ij...k}}
\end{align}
where $\prod_{(ij...k)}$ is the product over all the $d$-cells
$(ij...k)$.
If the above action amplitude  $\prod_{(ij...k)} [V_{ij...k}(\{g_{ij}\},\{g_i\})]^{s_{ij...k}}$ on closed  space-time complex ($\prt
M_\text{tri}=\emptyset $) is invariant under the gauge transformation
\begin{align}
\label{gaugeT}
 g_{ij} \to g'_{ij}=h_i g_{ij} h_j^{-1},\ \ g_i \to g'_i = h_ig_ih_i^{-1}\ \ \ h_i \in G_g
\end{align}
then the action amplitude $V_{ij...k}(\{g_{ij}\},\{g_i\})$ defines a gauge
theory of gauge group $G_g$.  If the  action amplitude is invariant under the
global transformation
\begin{align}
\label{symmT}
 g_{ij} \to g'_{ij}=h g_{ij} h^{-1},\ \ g_i \to g'_i = hg_i h^{-1}\ \ \ h \in G ,
\end{align}
then the action amplitude $V_{ij...k}(\{g_{ij}\},\{g_i\})$ defines a gauge
theory with a global symmetry $G_s=G/G_g$.  (We need to mod out $G_g$ since
when $h\in G_g$, it will generate a gauge transformation instead of a global
symmetry transformation.)

Using a cocycle $\nu_d( g_0, g_1,...,g_d) \in \cH^d(G,\R/\Z),\ g_i\in G$, we
can construct an  action amplitude $V_{ij...k}(\{g_{ij}\},\{g_i\})$ that define
a gauge theory with gauge group $G_s$ and global symmetry $G_s$.  First, we
note that the cocycle satisfies the cocycle condition
\begin{align}
&
\nu_d( g_0, g_1,...,g_d)=\nu_d( hg_0, hg_1,...,hg_d), \ \ \ h\in G
\nonumber\\
&
\prod_i \nu_d( g_0,..., \hat g_i ,...,g_{d+1})=1
\end{align}
where $g_0,..., \hat g_i ,...,g_{d+1}$ is the sequence $g_0,..., g_i
,...,g_{d+1}$ with $g_i$ removed.
The gauge theory action amplitude is given by
\begin{align}
 V_{01...d}(\{g_{ij}\},\{g_i\})&=0, \text{ if } g_{ij}g_{jk}\neq g_{ik}
\\
 V_{01...d}(\{g_{ij}\},\{g_i\})&=
\nu_d( \t g_0 g_0, \t g_1 g_1,..., \t g_d g_d), \text{ otherwise},
\nonumber 
\end{align}
where $\t g_i$ are given by
\begin{align}
 \t g_0 &=1, & \t g_1 &=\t g_0 g_{01},  
&
 \t g_2 &=\t g_1 g_{12}, & \t g_3 &=\t g_2 g_{23}, ...  
\end{align}
One can check that the above action amplitude $ V_{01...d}(\{g_{ij}\},\{g_i\})$ is invariant under the gauge transformation \eq{gaugeT}
and the global symmetry transformation
\eq{symmT}. Thus it defines an symmetric gauge theory

We know that the action amplitude is non-zero only when $g_{ij}g_{jk}= g_{ik}$.
The condition $g_{ij}g_{jk}= g_{ik}$ is the flat connection condition, and the
corresponding gauge theory is in the weak-coupling limit (actually is at the
zero-coupling).  This condition can be implemented precisely only when $G_g$ is
finite.  With  the flat connection condition  $g_{ij}g_{jk}= g_{ik}$, $\t
g_i$'s and the gauge equivalent sets of $g_{ij}$ have an one-to-one
correspondence.  

Since the total action amplitude  $\prod_{(ij...k)}
[V_{ij...k}(\{g_{ij}\},\{g_i\})]^{s_{ij...k}}$ on a sphere is always equal to 1
if the gauge flux vanishes, therefore $V_{ij...k}(\{g_{ij}\},\{g_i\})$
describes a quantized topological term in weak-coupling gauge theory (or
zero-coupling gauge theory).  This way, we show that \emph{quantized
topological term in a weak-coupling gauge theory with gauge group $G_g$ and
symmetry group $G_s$ can be constructed from each element of $\cH^d(G,\R/\Z)$}.

When $G_g=\{1\}$ (or $G=G_s$),
\begin{align}
 V_{01...d}(\{g_{ij}\},\{g_i\})&=
\nu_d( g_0, g_1,...,  g_d)
\end{align}
become the action amplitude for the topological non-linear $\si$-model,
describing the SPT phase labeled by the cocycle $\nu_d\in
\cH^d[G_s,\R/\Z)$.\cite{CGL1172}

When $G_s=\{1\}$ (or $G=G_g$),
\begin{align}
 V_{01...d}(\{g_{ij}\},\{g_i\})&=
\nu_d(\t g_0 g_0, \t g_1g_1,...,  \t g_dg_d).
\end{align}
We can use the gauge transformation \eq{gaugeT} to set $g_i=1$ in the above
and obtain
\begin{align}
 V_{01...d}(\{g_{ij}\},\{g_i\})&=
\nu_d(\t g_0 , \t g_1,...,  \t g_d).
\end{align}
This is the topological gauge theory studied in \Ref{DW9093,HW1267}.

\section{An approach based on classifying space}

In this section, we will consider the cases where $G_s,G_g$ can be finite or
continuous.  But at time being, we can only handle the situation where
$G=G_s\times G_g$.  Our approach is based on the classifying space.

\subsection{Motivations and results}

Let us first review some known results.  To gain a systematic understand of SRE
states with on-site symmetry $G_s$, we started with a non-linear $\si$-model
\begin{align}
 \cL= \frac{1}{\la_s} [\prt g(x^\mu)]^2, \ \ g\in G_s
\end{align}
with symmetry group $G_s$ as the target space.  The model can be in a
disordered phase that does not break the symmetry $G_s$ when $\la$ is large.
By adding different $2\pi$ quantized topological $\th$-terms to the Lagrangian
$\cL$, we can get different Lagrangians that describe different disordered
phases that does not break the symmetry $G_s$.\cite{CGL1172}  Those disordered
phases are the symmetry protected topological (SPT)
phases.\cite{GW0931,PBT0959} So we can use the quantized topological terms to
classify the SPT phases.  (In general, topological terms, by definition, are
the terms that do not depend on space-time metrics.)

We know that gauge theory
\begin{align}
 \cL= \frac{1}{\la}\Tr(F_{\mu\nu}F^{\mu\nu})
\end{align}
is one way to describe LRE states (\ie topologically ordered states).  In
\Ref{DW9093,HW1267}, different quantized topological terms in weak-coupling
gauge theory with gauge group $G_g$ and small $\la$ in $d$ space-time
dimensions are constructed and classified, using the topological cohomology
class $H^{d+1}(BG_g,\Z)$ for the classifying space $BG_g$ of the gauge group
$G_g$.  By adding those quantized topological terms to the above Lagrangian for
the weak-coupling gauge theory, we may obtain different phases of the
weak-coupling gauge theory.

In this section, we plan to combine the above two approaches by studying the
quantized topological terms in the combined theory
\begin{align}
\label{swgauge}
 \cL= \frac{1}{\la}\Tr(F_{\mu\nu}F^{\mu\nu})+ \frac{1}{\la_s} [\prt g(x^\mu)]^2, \ \ g\in G_s
\end{align}
where $F$ is the field strength with gauge group $G_g$, and $(\la,\la_s) \to $
(small,large).  Such a theory is a gauge theory with symmetry $G_s$.  We find
that  quantized topological terms in the combined theory can be constructed and
classified by the topological cohomology class $H^{d+1}(BG_s\times BG_g,\Z)$ for
the classifying space of the product $G_g\times G_s$.  Those quantized
topological terms give us a somewhat systematic understanding of different
phases of weak coupling gauge theories with symmetry.  If those  symmetric weak
coupling gauge theories are gapped (for example, for finite gauge groups), then
the theories will describe topologically ordered states with symmetry. Those
SET phases in $d$ space-time dimensions are described by elements in
$H^{d+1}(BG_s\times BG_g,\Z)$.

\subsection{Gauge theory as a non-linear $\si$-model with classifying
space as the target space}
\label{top_sigmamodel}

To obtain the above result, we will follow closely the approaches used in
\Ref{HW1267} and \Ref{CGL1172}. We will obtained our result in two steps.

\subsubsection{Symmetric weak-coupling gauge theory as the
non-linear $\si$-model of $G_s\times BG_g$}

As in \Ref{HW1267}, we may view a  weak-coupling gauge theory with gauge group
$G_g$ as a non-linear $\si$-model with classifying space $BG_g$ as the target
space.  So the symmetric weak-coupling gauge theory in \eqn{swgauge} can be
viewed as  a non-linear $\si$-model with $G_s\times BG_g$ as the target space,
where each path in the path integral is given by an \emph{embedding} $\ga:
M_\text{tri} \to G_s\times BG_g$ from the space-time complex $ M_\text{tri}$ to
$G_s\times BG_g$.  We can study topological terms in our symmetric
weak-coupling gauge theory by studying the topological terms in the
corresponding non-linear $\si$-model.

Following \Ref{HW1267}, a total term $S_\text{top}$ corresponds to evaluating a
cocycle $\al_{d} \in Z(G_s\times BG_g,\R/\Z)$ on the complex $\ga(M_\text{tri})
\subset G_s\times BG_g$:
\begin{equation}
\label{topgauge}
S_\text{top}[\ga]=2\pi \< \al_{d}, \ga(M_\text{tri})\> \ \ \text{mod } 2\pi.
\end{equation}
Such a topological term does not depend on any smooth deformation of $\ga$ and
is thus ``topological''.  (Note that the evaluation of the  $d$-cocycle on any
$d$-cycles [\ie $d$-dimensional closed complexes] are equal to $0$ mod 1 if the
$d$-cycles are boundaries of some $(d+1)$-dimensional complex.)

Here we would like to stress that the cocycle
$\al_{d}$ on the group manifold is \emph{not} the ordinary topological
cocycle. It has a symmetry condition
\begin{align}
 \< \al_{d}, c\>=
 \< \al_{d}, c_g\>
\end{align}
where $c$ is a complex in $G_s$, and $c_g$ is the complex generated from $c$
by the symmetry transformation $G_s \to gG_s$, $g\in G_s$.
Also, since $\la_s\to \infty$ and $g(x^\mu)$ have large fluctuations in \eqn{swgauge}, $\< \al_{d}, c\>$ only depend on the vertices $g_0,g_1,...$ of $c$:
\begin{align}
 \< \al_{d}, c\> &=\nu(g_0,g_1,...),\ \ \ \
 \nu(gg_0,gg_1,...) =
 \nu(g_0,g_1,...);
\nonumber\\
g,g_i & \in G_s.
\end{align}
So, on $G_s$, $\al_d$ is actually a cocycle in the group cohomology
$\cZ(G_s,\R/\Z)$,\cite{CGL1172} while on $BG_g$, $\al_d$ is the usual cocycle in
the topological cohomology $Z(BG_g,\R/\Z)$.

Since, on $G_s$, $\al_d$ is a cocycle in the group cohomology $\cZ(G_s,\R/\Z)$,
when $G_s$ contain anti-unitary symmetry, such anti-unitary symmetry
transformation will have a non-trivial action on $\R/\Z$: $x \to -x$, $x\in
\R/\Z$.\cite{CGL1172}

If two $d$-cocycles, $\al_{d}, \al'_{d} \in Z^{d}(BG_g,\R/\Z)$, differ by a
coboundary: $\al'_{d} - \al_{d} = \dd \mu_d$, $\mu_d \in C^d(BG_g,\R/\Z)$, then,
the corresponding action amplitudes, $\e^{\imth S_\text{top}[\ga]}$ and
$\e^{\imth S_\text{top}'[\ga]}$, can smoothly deform into each other without
phase transition.  So $\e^{\imth S_\text{top}[\ga]}$ and $\e^{\imth
S_\text{top}'[\ga]}$, or $\al_{d}$ and $\al'_{d}$, describe the same quantum
phase.  Therefore, we regard $\al_{d}$ and $\al'_{d}$ to be equivalent.  The
equivalent classes of the $d$-cocycles form the $d$ cohomology class
$H^{d}(G_s\times BG_g,\R/\Z)$.  We conclude that the  topological terms in
symmetric weak-coupling lattice gauge theories are described by
$H^{d}(G_s\times BG_g,\R/\Z)$ in $d$ space-time dimensions.

To calculate
$H^{d}(G_s\times BG_g,\R/\Z)$, let us first calculate
$H^{d}(G_s\times BG_g,\Z)$. Using the
the K\"unneth formula \eqn{kunnZ} (with $M'=\Z$), we find that
\begin{align}
&\ \ \ \ H^d(G_s\times BG_g,\Z)
\nonumber\\
&\simeq \Big[\oplus_{p=0}^d \cH^p(G_s,\Z)\otimes_{\Z} H^{d-p}(BG_g,\Z)\Big]\oplus
\nonumber\\
&\ \ \ \ \ \
\Big[\oplus_{p=0}^{d+1}
\text{Tor}_1^{\Z}[\cH^p(G_s,{\Z}),H^{d-p+1}(BG_g,\Z)]\Big]  .
\end{align}
In the above, we have used the fact that the cohomology on $G_s$ is the group
cohomology $\cH$ and the cohomology on $BG_g$ is the usual topological cohomology
$H$.

In appendix \ref{HBGRZ}, we show that (see \eqn{ucf})
\begin{align}
\label{ucfU}
&\ \ \ \ H^d(X,\R/\Z)
\\
&\simeq  H^d(X,\Z)\otimes_{\Z} \R/\Z \oplus
\text{Tor}_1^{\Z}[H^{d+1}(X,{\Z}),\R/\Z]  .
\nonumber
\end{align}
Using
\begin{align}
\Z \otimes_{\Z} \R/\Z =\R/\Z, \ \ \ \
\Z_n \otimes_{\Z} \R/\Z =0,
\nonumber\\
\text{Tor}_1^{\Z}(\Z,\R/\Z)=0, \ \ \ \
\text{Tor}_1^{\Z}(\Z_n,\R/\Z)=\Z_n,
\end{align}
we see that
$H^d (X,\R/\Z)$ has a form $H^d (X,\R/\Z)=\R/\Z\oplus ... \oplus
\R/\Z \oplus Z_{n_1} \oplus Z_{n_2}\oplus ...  $.
So the discrete part of
$H^d (X,\R/\Z)$ is given by
\begin{align}
 \text{Dis}[H^d (X,\R/\Z)]&=Z_{n_1} \oplus Z_{n_2}\oplus ...
\nonumber\\
&= \text{Tor}[H^{d+1}(X,\Z)] ,
\end{align}
where we have used
\begin{align}
H^{d+1}(X,\Z) = \text{Free}[H^{d+1}(X,\Z)]\oplus \text{Tor}[H^{d+1}(X,\Z)]
\end{align}
with $\text{Tor}[H^{d+1}(X,\Z)]$ the torsion part and
$\text{Free}[H^{d+1}(X,\Z)]$ the free part of $H^{d+1}(X,\Z)$.
Therefore, we have
\begin{align}
&\ \ \ \ \text{Dis}[H^d(G_s\times BG_g,\R/\Z)]
\nonumber\\
&\simeq \text{Tor}
\Big[[\oplus_{p=0}^{d+1} \cH^p(G_s,\Z)\otimes_{\Z} H^{d+1-p}(BG_g,\Z)]\oplus
\nonumber\\
&\ \ \ \ \ \
[\oplus_{p=0}^{d+2}
\text{Tor}_1^{\Z}(\cH^p(G_s,{\Z}),H^{d-p+2}(BG_g,\Z))]\Big]  .
\end{align}
Since $\cH^d(G_s,\Z)=H^d(BG_s,\Z)$, the above can be rewritten as
\begin{align}
\label{res2}
&\ \ \ \ \text{Dis}[H^d(G_s\times BG_g,\R/\Z)]
\nonumber\\
&\simeq \text{Tor}
\Big[[\oplus_{p=0}^{d+1} H^p(BG_s,\Z)\otimes_{\Z} H^{d+1-p}(BG_g,\Z)]\oplus
\nonumber\\
&\ \ \ \ \ \
[\oplus_{p=0}^{d+2}
\text{Tor}_1^{\Z}(H^p(BG_s,{\Z}),H^{d-p+2}(BG_g,\Z))]\Big]
\nonumber\\
& =
\Big[ \oplus_{p=0}^{d+1} \text{Tor}[H^p(BG_s,\Z)]\otimes_{\Z} \text{Tor}[H^{d+1-p}(BG_g,\Z)]
\Big]\oplus
\nonumber\\
&\ \ \ \
\Big[ \oplus_{p=0}^{d+1} \text{Free}[H^p(BG_s,\Z)]\otimes_{\Z} \text{Tor}[H^{d+1-p}(BG_g,\Z)]
\Big]\oplus
\nonumber\\
&\ \ \ \
\Big[ \oplus_{p=0}^{d+1} \text{Tor}[H^p(BG_s,\Z)]\otimes_{\Z} \text{Free}[H^{d+1-p}(BG_g,\Z)]
\Big]\oplus
\nonumber\\
&\ \ \ \
\Big[\oplus_{p=0}^{d+2}
\text{Tor}_1^{\Z}(H^p(BG_s,{\Z}),H^{d-p+2}(BG_g,\Z))\Big]
.
\end{align}
Each element in the above cohomology class
describes a quantized topological term in the weakly coupled gauge theory
with symmetry $G_s$.

\subsubsection{Chern-Simons form}

We note that
\begin{align}
&\ \ \ \ H^{d+1}(BG_s\times BG_g,\Z)
\nonumber\\
& =
\Big[ \oplus_{p=0}^{d+1} H^p(BG_s,\Z)\otimes_{\Z} H^{d+1-p}(BG_g,\Z)
\Big]\oplus
\nonumber\\
&\ \ \ \ \ \
\Big[\oplus_{p=0}^{d+2}
\text{Tor}_1^{\Z}[H^p(BG_s,{\Z}),H^{d-p+2}(BG_g,\Z)]\Big]
.
\end{align}
So the result \eq{res2} is very close to our proposal that elements in
$H^{d+1}(BG_s\times BG_g,\Z)$ correspond to the quantized topological terms.
The only thing missing is the free part of $H^{d}(BG_g,\Z)$.

In fact, the free part of  $H^{d+1}(BG_g,\Z)$, denoted as
Free$[H^{d+1}(BG_g,\Z)]$, is non-zero only when $d=$odd. So in the following,
we will consider only $d$= odd cases.  The free part Free$[H^{d+1}(BG_g,\Z)]$
corresponds to the Chern-Simons forms in $d$ space-time dimensions.

To understand such a result, we first choose a
$\om \in \text{Free}[H^{d+1}(BG_g,\Z)]$.  We can find integers $K_i$ such that
\be
- \om
+\frac{K_1}{\frac{d+1}{2}!(2\pi)^{\frac{d+1}{2}}}\Tr F^{\frac{d+1}{2}}
+ \cdots
\ee
is an exact form $\dd \th_d(A)$.
Here $\th_d(A)$ is called a Chern-Simons form in $d$-dimensions.

We can use a Chern-Simons form $\th_{d-p}(A)$ and a cocycle
$\al_p \in \cH^p(G_s,\Z)$
to construct a quantized topological term
\begin{align}
S_\text{top}[\ga]=2\pi \< \al_{p}\cup \th_{d-p}(A), \ga(M_\text{tri})\> \ \ \text{mod } 2\pi.
\end{align}
Such kind of topological terms are labeled by the elements in
\begin{align}
&\ \ \ \
\oplus_{p=0}^{d+1} \cH^p(G_s,\Z)\otimes_{\Z} \text{Free}[H^{d+1-p}(BG_g,\Z)]
\nonumber\\
&=
\oplus_{p=0}^{d+1} H^p(BG_s,\Z)\otimes_{\Z} \text{Free}[H^{d+1-p}(BG_g,\Z)].
\end{align}
Combining the above result with \eqn{res2}, we find that the elements in
$H^{d+1}(BG_s\times BG_g,\Z)$ correspond to the quantized topological terms.

\section{An example: $G_s=Z_2$ and $G_g=Z_2$}

In this section, we will discuss a simple example with $G_s=Z_2$ and $G_g=Z_2$.
There are two kinds of extensions $G$ of $G_s=Z_2$ by $G_g=Z_2$: $G=Z_2\times
Z_2$ and $G=Z_4$.  So the quantized topological terms and the SET phases are
described by $\cH^d(Z_2\times Z_2,\R/\Z)$ and $\cH^d(Z_4,\R/\Z)$ in $d$
space-time dimensions.

In $d=3$ space-time dimensions, we have
\begin{align}
 \cH^3(Z_2\times Z_2,\R/\Z) = \Z_2^3,\ \ \
 \cH^3(Z_4,\R/\Z) = \Z_4.
\end{align}
So there are 12 SET phases for weak-coupling $Z_2$ gauge theory with $Z_2$
symmetry. However, at this stage, it is not clear if those 12 SET phases are
really distinct, since they could be smoothly connected via strong coupling
gauge theory. Later, we will see that the  12 SET phases are indeed distinct,
since they have distinct physical properties.

\subsection{A $K$-matrix approach}

To understand the physical properties of those 12 SET phases,
we would like to use Levin-Gu duality to gauge the $G_s$
and turn the theory into gauge theory with gauge group $G$.

Let us first consider the $G=Z_2\times Z_2$ case.
A $G=Z_2\times Z_2$ gauge theory can be described by
$U^4(1)$ mutual Chern-Simons theory:\cite{HOS0497,KLW0834}
\begin{align}
 \cL=\frac{1}{4\pi} K_{0,IJ} a^I_\mu\prt_\nu a^J_\la + ...
\end{align}
with
\begin{align}
 K_0=2\bpm
0&1&0&0\\
1&0&0&0\\
0&0&0&1\\
0&0&1&0\\
\epm
\end{align}
The $G_s$ charge corresponds to the unit charge of $a^1_\mu$ gauge field and
the $G_g$ gauge charge corresponds to the unit charge of $a^3_\mu$ gauge field.
The $G_s$ flux excitation (in the $G=Z_2\times Z_2$ gauge theory) corresponds
to the end of branch-cut in the original theory along which we have a twist
generated by a $G_s$ symmetry transformation (see \Ref{LG1220} for a detailed
discussion about the symmetry twist).  Such  $G_s$-flux correspond to the flux
of $a^1_\mu$ gauge field.

The 8 types of quantized topological terms are given by
\begin{align}
 W_\text{top}=
\frac{n_1}{2\pi}  a^1_\mu\prt_\nu a^1_\la +
\frac{n_{12}}{2\pi}  a^1_\mu\prt_\nu a^3_\la +
\frac{n_2}{2\pi}  a^3_\mu\prt_\nu a^3_\la
\end{align}
$n_1=0,1$, $n_{12}=0,1$, $n_2=0,1$.
The total Lagrangian has a form
\begin{align}
 \cL+W_\text{top}=\frac{1}{4\pi} K_{IJ} a^I_\mu\prt_\nu a^J_\la + ...
\end{align}
with
\begin{align}
\label{csz2z2}
 K=\bpm
2n_1&2&n_{12}&0\\
2&0&0&0\\
n_{12}&0&2n_2&2\\
0&0&2&0\\
\epm .
\end{align}
Two $K$-matrices are equivalent: $K_1\sim K_2$ if $K_1=U^TK_2U$
for an integer matrix with det$(U)=\pm 1$. We find
$K(n_1,n_{12},n_2) \sim K(n_1+2,n_{12},n_2) \sim K(n_1,n_{12}+2,n_2) \sim K(n_1,n_{12},n_2+2)$.
Thus only $n_1,n_{12},n_2=0,1$ give rise to inequivalent $K$-matrices.

A particle carrying $l_I$ $a^I_\mu$-charge will have a statistics
\begin{align}
 \th_l=\pi l_I (K^{-1})^{IJ}l_J .
\end{align}
A particle carrying $l_I$ $a^I_\mu$-charge will have a mutual statistics
with a particle carrying $\t l_I$ $a^I_\mu$-charge:
\begin{align} \label{mutual}
 \th_{l,\t l}=2\pi l_I (K^{-1})^{IJ}\t l_J .
\end{align}

We note that \emph{the $G_s$ charge is identified with the unit
$a^1_\mu$-charge} and the $G_g$ gauge charge is identified with the unit
$a^3_\mu$-charge.  Using
\begin{align}
 K^{-1}=\frac{1}{4}\bpm
0&2&0&0\\
2&-2n_1&0&-n_{12}\\
0&0&0&2\\
0&-n_{12}&2&-2n_2\\
\epm ,
\end{align}
we find that the $G_s$ charge (the unit $a^1_\mu$-charge) and the $G_g$ gauge
charge (the unit $a^3_\mu$-charge) remain bosonic after inclusion of the
topological terms.  This is actually a condition on the topological terms: the
topological terms do not affect the statistics of the gauge charge.

\begin{table}[tb]
\centering
 \begin{tabular}{ |c||c|c|c|c| }
 \hline
 \multicolumn{5}{|c|}{$(n_1n_{12}n_2)= (000) $} \\ 
 \hline
$(l_1l_2l_3l_4)$ & $G_s$-charge & $G_s$-twist & $G_g$-gauge & statistics \\
 \hline
\hline
(0000) & 0 & 0 & 0 & 0\\
(1000) & 1 & 0 & 0 & 0\\
(0010) & 0 & 0 & e & 0\\
(1010) & 1 & 0 & e & 0\\
(0001) & 0 & 0 & m & 0\\
(1001) & 1 & 0 & m & 0\\
(0011) & 0 & 0 & em & 1\\
(1011) & 1 & 0 & em & 1\\
\hline
(0100) & 0 & 1 & 0 & 0\\
(1100) & 1 & 1 & 0 & 1\\
(0110) & 0 & 1 & e & 0\\
(1110) & 1 & 1 & e & 1\\
(0101) & 0 & 1 & m & 0\\
(1101) & 1 & 1 & m & 1\\
(0111) & 0 & 1 & em & 1\\
(1111) & 1 & 1 & em & 0\\
\hline
 \end{tabular}
\caption{The  $G_s$-charges, the  $G_s$-twists, the  $G_g$-gauge sectors,
 and the statistics of the 16 kinds of quasiparticles/defects
 in the SET state $(n_1n_{12}n_2)$.
}
\label{Z2tb1}
\end{table}

\begin{table}[tb]
\centering
 \begin{tabular}{ |c||c|c|c|c| }
 \hline
 \multicolumn{5}{|c|}{$(n_1n_{12}n_2)= (010) $} \\ 
 \hline
$(l_1l_2l_3l_4)$ & $G_s$-charge & $G_s$-twist & $G_g$-gauge & statistics \\
 \hline
\hline
(0000) & 0 & 0 & 0 & 0\\
(1000) & 1 & 0 & 0 & 0\\
(0010) & 0 & 0 & e & 0\\
(1010) & 1 & 0 & e & 0\\
(0001) & $-1/2$ & 0 & m & 0\\
(1001) & $1/2$ & 0 & m & 0\\
(0011) & $-1/2$ & 0 & em & 1\\
(1011) & $1/2$ & 0 & em & 1\\
\hline
(0100) & 0 & 1 & 0 & 0\\
(1100) & 1 & 1 & 0 & 1\\
(0110) & 0 & 1 & e & 0\\
(1110) & 1 & 1 & e & 1\\
(0101) & $-1/2$ & 1 & m & $-1/2$\\
(1101) & $1/2$ & 1 & m & $1/2$\\
(0111) & $-1/2$ & 1 & em & $1/2$\\
(1111) & $1/2$ & 1 & em & $-1/2$\\
\hline
 \end{tabular}
\caption{The  $G_s$-charges, the  $G_s$-twists, the  $G_g$-gauge sectors,
 and the statistics of the 16 kinds of quasiparticles/defects
 in the SET state $(n_1n_{12}n_2)$.
}
\label{Z2tb2}
\end{table}

\begin{table}[tb]
\centering
 \begin{tabular}{ |c||c|c|c|c| }
 \hline
 \multicolumn{5}{|c|}{$(n_1n_{12}n_2)= (100) $} \\ 
 \hline
$(l_1l_2l_3l_4)$ & $G_s$-charge & $G_s$-twist & $G_g$-gauge & statistics \\
 \hline
\hline
(0000) & 0 & 0 & 0 & 0\\
(1000) & 1 & 0 & 0 & 0\\
(0010) & 0 & 0 & e & 0\\
(1010) & 1 & 0 & e & 0\\
(0001) & 0 & 0 & m & 0\\
(1001) & 1 & 0 & m & 0\\
(0011) & 0 & 0 & em & 1\\
(1011) & 1 & 0 & em & 1\\
\hline
(0100) & $-1/2$ & 1 & 0 & $-1/2$\\
(1100) & $1/2$ & 1 & 0 & $1/2$\\
(0110) & $-1/2$ & 1 & e & $-1/2$\\
(1110) & $1/2$ & 1 & e & $1/2$\\
(0101) & $-1/2$ & 1 & m & $-1/2$\\
(1101) & $1/2$ & 1 & m & $1/2$\\
(0111) & $-1/2$ & 1 & em & $1/2$\\
(1111) & $1/2$ & 1 & em & $-1/2$\\
\hline
 \end{tabular}
\caption{The  $G_s$-charges, the  $G_s$-twists, the  $G_g$-gauge sectors,
 and the statistics of the 16 kinds of quasiparticles/defects
 in the SET state $(n_1n_{12}n_2)$.
}
\label{Z2tb3}
\end{table}

\begin{table}[tb]
\centering
 \begin{tabular}{ |c||c|c|c|c| }
 \hline
 \multicolumn{5}{|c|}{$(n_1n_{12}n_2)= (110) $} \\ 
 \hline
$(l_1l_2l_3l_4)$ & $G_s$-charge & $G_s$-twist & $G_g$-gauge & statistics \\
 \hline
\hline
(0000) & 0 & 0 & 0 & 0\\
(1000) & 1 & 0 & 0 & 0\\
(0010) & 0 & 0 & e & 0\\
(1010) & 1 & 0 & e & 0\\
(0001) & $-1/2$ & 0 & m & 0\\
(1001) & $1/2$ & 0 & m & 0\\
(0011) & $-1/2$ & 0 & em & 1\\
(1011) & $1/2$ & 0 & em & 1\\
\hline
(0100) & $-1/2$ & 1 & 0 & $-1/2$\\
(1100) & $1/2$ & 1 & 0 & $1/2$\\
(0110) & $-1/2$ & 1 & e & $-1/2$\\
(1110) & $1/2$ & 1 & e & $1/2$\\
(0101) & 1 & 1 & m & 1\\
(1101) & 0 & 1 & m & 0\\
(0111) & 1 & 1 & em & 0\\
(1111) & 0 & 1 & em & 1\\
\hline
 \end{tabular}
\caption{The  $G_s$-charges, the  $G_s$-twists, the  $G_g$-gauge sectors,
 and the statistics of the 16 kinds of quasiparticles/defects
 in the SET state $(n_1n_{12}n_2)$.
}
\label{Z2tb4}
\end{table}

\begin{table}[tb]
\centering
 \begin{tabular}{ |c||c|c|c|c| }
 \hline
 \multicolumn{5}{|c|}{$(n_1n_{12}n_2)= (001) $} \\ 
 \hline
$(l_1l_2l_3l_4)$ & $G_s$-charge & $G_s$-twist & $G_g$-gauge & statistics \\
 \hline
\hline
(0000) & 0 & 0 & 0 & 0\\
(1000) & 1 & 0 & 0 & 0\\
(0010) & 0 & 0 & e & 0\\
(1010) & 1 & 0 & e & 0\\
(0001) & 0 & 0 & m & $-1/2$\\
(1001) & 1 & 0 & m & $-1/2$\\
(0011) & 0 & 0 & em & $1/2$\\
(1011) & 1 & 0 & em & $1/2$\\
\hline
(0100) & 0 & 1 & 0 & 0\\
(1100) & 1 & 1 & 0 & 1\\
(0110) & 0 & 1 & e & 0\\
(1110) & 1 & 1 & e & 1\\
(0101) & 0 & 1 & m & $-1/2$\\
(1101) & 1 & 1 & m & $1/2$\\
(0111) & 0 & 1 & em & $1/2$\\
(1111) & 1 & 1 & em & $-1/2$\\
\hline
 \end{tabular}
\caption{The  $G_s$-charges, the  $G_s$-twists, the  $G_g$-gauge sectors,
 and the statistics of the 16 kinds of quasiparticles/defects
 in the SET state $(n_1n_{12}n_2)$.
}
\label{Z2tb5}
\end{table}

\begin{table}[tb]
\centering
 \begin{tabular}{ |c||c|c|c|c| }
 \hline
 \multicolumn{5}{|c|}{$(n_1n_{12}n_2)= (011) $} \\ 
 \hline
$(l_1l_2l_3l_4)$ & $G_s$-charge & $G_s$-twist & $G_g$-gauge & statistics \\
 \hline
\hline
(0000) & 0 & 0 & 0 & 0\\
(1000) & 1 & 0 & 0 & 0\\
(0010) & 0 & 0 & e & 0\\
(1010) & 1 & 0 & e & 0\\
(0001) & $-1/2$ & 0 & m & $-1/2$\\
(1001) & $1/2$ & 0 & m & $-1/2$\\
(0011) & $-1/2$ & 0 & em & $1/2$\\
(1011) & $1/2$ & 0 & em & $1/2$\\
\hline
(0100) & 0 & 1 & 0 & 0\\
(1100) & 1 & 1 & 0 & 1\\
(0110) & 0 & 1 & e & 0\\
(1110) & 1 & 1 & e & 1\\
(0101) & $-1/2$ & 1 & m & 1\\
(1101) & $1/2$ & 1 & m & 0\\
(0111) & $-1/2$ & 1 & em & 0\\
(1111) & $1/2$ & 1 & em & 1\\
\hline
 \end{tabular}
\caption{The  $G_s$-charges, the  $G_s$-twists, the  $G_g$-gauge sectors,
 and the statistics of the 16 kinds of quasiparticles/defects
 in the SET state $(n_1n_{12}n_2)$.
}
\label{Z2tb6}
\end{table}

\begin{table}[tb]
\centering
 \begin{tabular}{ |c||c|c|c|c| }
 \hline
 \multicolumn{5}{|c|}{$(n_1n_{12}n_2)= (101) $} \\ 
 \hline
$(l_1l_2l_3l_4)$ & $G_s$-charge & $G_s$-twist & $G_g$-gauge & statistics \\
 \hline
\hline
(0000) & 0 & 0 & 0 & 0\\
(1000) & 1 & 0 & 0 & 0\\
(0010) & 0 & 0 & e & 0\\
(1010) & 1 & 0 & e & 0\\
(0001) & 0 & 0 & m & $-1/2$\\
(1001) & 1 & 0 & m & $-1/2$\\
(0011) & 0 & 0 & em & $1/2$\\
(1011) & 1 & 0 & em & $1/2$\\
\hline
(0100) & $-1/2$ & 1 & 0 & $-1/2$\\
(1100) & $1/2$ & 1 & 0 & $1/2$\\
(0110) & $-1/2$ & 1 & e & $-1/2$\\
(1110) & $1/2$ & 1 & e & $1/2$\\
(0101) & $-1/2$ & 1 & m & 1\\
(1101) & $1/2$ & 1 & m & 0\\
(0111) & $-1/2$ & 1 & em & 0\\
(1111) & $1/2$ & 1 & em & 1\\
\hline
 \end{tabular}
\caption{The  $G_s$-charges, the  $G_s$-twists, the  $G_g$-gauge sectors,
 and the statistics of the 16 kinds of quasiparticles/defects
 in the SET state $(n_1n_{12}n_2)$.
}
\label{Z2tb7}
\end{table}

\begin{table}[tb]
\centering
 \begin{tabular}{ |c||c|c|c|c| }
 \hline
 \multicolumn{5}{|c|}{$(n_1n_{12}n_2)= (111) $} \\ 
 \hline
$(l_1l_2l_3l_4)$ & $G_s$-charge & $G_s$-twist & $G_g$-gauge & statistics \\
 \hline
\hline
(0000) & 0 & 0 & 0 & 0\\
(1000) & 1 & 0 & 0 & 0\\
(0010) & 0 & 0 & e & 0\\
(1010) & 1 & 0 & e & 0\\
(0001) & $-1/2$ & 0 & m & $-1/2$\\
(1001) & $1/2$ & 0 & m & $-1/2$\\
(0011) & $-1/2$ & 0 & em & $1/2$\\
(1011) & $1/2$ & 0 & em & $1/2$\\
\hline
(0100) & $-1/2$ & 1 & 0 & $-1/2$\\
(1100) & $1/2$ & 1 & 0 & $1/2$\\
(0110) & $-1/2$ & 1 & e & $-1/2$\\
(1110) & $1/2$ & 1 & e & $1/2$\\
(0101) & 1 & 1 & m & $1/2$\\
(1101) & 0 & 1 & m & $-1/2$\\
(0111) & 1 & 1 & em & $-1/2$\\
(1111) & 0 & 1 & em & $1/2$\\
\hline
 \end{tabular}
\caption{The  $G_s$-charges, the  $G_s$-twists, the  $G_g$-gauge sectors,
 and the statistics of the 16 kinds of quasiparticles/defects
 in the SET state $(n_1n_{12}n_2)$.
}
\label{Z2tb8}
\end{table}

\begin{table}[tb]
\centering
 \begin{tabular}{ |c||c|c|c|c| }
 \hline
 \multicolumn{5}{|c|}{$m_1=0$} \\ 
 \hline
$(l_1l_2)$ & $G_s$-charge & $G_s$-twist & $G_g$-gauge & statistics \\
 \hline
(00) & 0 & 0 & 0 & 0\\
(20) & 1 & 0 & 0 & 0\\
(10) & $1/2$ & 0 & e & 0\\
(30) & $-1/2$ & 0 & e & 0\\
(02) & 0 & 0 & m & 0\\
(22) & 1 & 0 & m & 0\\
(12) & $1/2$ & 0 & em & 1\\
(32) & $-1/2$ & 0 & em & 1\\
 \hline
(01) & 0 & 1 & 0 & 0\\
(21) & 1 & 1 & 0 & 1\\
(11) & $1/2$ & 1 & e & $1/2$\\
(31) & $-1/2$ & 1 & e & $-1/2$\\
(03) & 0 & 1 & m & 0\\
(23) & 1 & 1 & m & 1\\
(13) & $1/2$ & 1 & em & $-1/2$\\
(33) & $-1/2$ & 1 & em & $1/2$\\
\hline
 \end{tabular}
\caption{
 The  $G_s$-charges, the  $G_s$-twists,
 the $G_g$-gauge sectors, and the statistics of the 16 kinds of
 quasiparticles/defects in the SET state $m_1=0$ with $\v q^T =(1/2,-m_1/4)$.
}
\label{Z4tb1}
\end{table}

\begin{table}[tb]
\centering
 \begin{tabular}{ |c||c|c|c|c| }
 \hline
 \multicolumn{5}{|c|}{$m_1=1$} \\ 
 \hline
$(l_1l_2)$ & $G_s$-charge & $G_s$-twist & $G_g$-gauge & statistics \\
 \hline
(00) & 0 & 0 & 0 & 0\\
(20) & 1 & 0 & 0 & 0\\
(10) & $1/2$ & 0 & e & 0\\
(30) & $-1/2$ & 0 & e & 0\\
(02) & $-1/2$ & 0 & m & $-1/2$\\
(22) & $1/2$ & 0 & m & $-1/2$\\
(12) & 0 & 0 & em & $1/2$\\
(32) & 1 & 0 & em & $1/2$\\
 \hline
(01) & $-1/4$ & 1 & 0 & $-1/8$\\
(21) & $3/4$ & 1 & 0 & $7/8$\\
(11) & $1/4$ & 1 & e & $3/8$\\
(31) & $-3/4$ & 1 & e & $-5/8$\\
(03) & $-3/4$ & 1 & m & $7/8$\\
(23) & $1/4$ & 1 & m & $-1/8$\\
(13) & $-1/4$ & 1 & em & $3/8$\\
(33) & $3/4$ & 1 & em & $-5/8$\\
\hline
 \end{tabular}
\caption{
 The  $G_s$-charges, the  $G_s$-twists,
 the $G_g$-gauge sectors, and the statistics of the 16 kinds of
 quasiparticles/defects in the SET state $m_1=1$ with $\v q^T =(1/2,-m_1/4)$.
}
\label{Z4tb2}
\end{table}

\begin{table}[tb]
\centering
 \begin{tabular}{ |c||c|c|c|c| }
 \hline
 \multicolumn{5}{|c|}{$m_1=2$} \\ 
 \hline
$(l_1l_2)$ & $G_s$-charge & $G_s$-twist & $G_g$-gauge & statistics \\
 \hline
(00) & 0 & 0 & 0 & 0\\
(20) & 1 & 0 & 0 & 0\\
(10) & $1/2$ & 0 & e & 0\\
(30) & $-1/2$ & 0 & e & 0\\
(02) & 1 & 0 & em & 1\\
(22) & 0 & 0 & em & 1\\
(12) & $-1/2$ & 0 & m & 0\\
(32) & $1/2$ & 0 & m & 0\\
 \hline
(01) & $-1/2$ & 1 & 0 & $-1/4$\\
(21) & $1/2$ & 1 & 0 & $3/4$\\
(11) & 0 & 1 & e & $1/4$\\
(31) & 1 & 1 & e & $-3/4$\\
(03) & $1/2$ & 1 & em & $-1/4$\\
(23) & $-1/2$ & 1 & em & $3/4$\\
(13) & 1 & 1 & m & $-3/4$\\
(33) & 0 & 1 & m & $1/4$\\
\hline
 \end{tabular}
\caption{
 The  $G_s$-charges, the  $G_s$-twists,
 the $G_g$-gauge sectors, and the statistics of the 16 kinds of
 quasiparticles/defects in the SET state $m_1=2$ with $\v q^T =(1/2,-m_1/4)$.
}
\label{Z4tb3}
\end{table}

\begin{table}[tb]
\centering
 \begin{tabular}{ |c||c|c|c|c| }
 \hline
 \multicolumn{5}{|c|}{$m_1=3$} \\ 
 \hline
$(l_1l_2)$ & $G_s$-charge & $G_s$-twist & $G_g$-gauge & statistics \\
 \hline
(00) & 0 & 0 & 0 & 0\\
(20) & 1 & 0 & 0 & 0\\
(10) & $1/2$ & 0 & e & 0\\
(30) & $-1/2$ & 0 & e & 0\\
(02) & $1/2$ & 0 & m & $1/2$\\
(22) & $-1/2$ & 0 & m & $1/2$\\
(12) & 1 & 0 & em & $-1/2$\\
(32) & 0 & 0 & em & $-1/2$\\
 \hline
(01) & $-3/4$ & 1 & 0 & $-3/8$\\
(21) & $1/4$ & 1 & 0 & $5/8$\\
(11) & $-1/4$ & 1 & e & $1/8$\\
(31) & $3/4$ & 1 & e & $-7/8$\\
(03) & $-1/4$ & 1 & m & $5/8$\\
(23) & $3/4$ & 1 & m & $-3/8$\\
(13) & $1/4$ & 1 & em & $1/8$\\
(33) & $-3/4$ & 1 & em & $-7/8$\\
\hline
 \end{tabular}
\caption{
 The  $G_s$-charges, the  $G_s$-twists,
 the $G_g$-gauge sectors, and the statistics of the 16 kinds of
 quasiparticles/defects in the SET state $m_1=3$ with $\v q^T =(1/2,-m_1/4)$.
}
\label{Z4tb4}
\end{table}

\emph{The end of branch-cut in the original theory correspond to $\pi$-flux in
$a^1_\mu$}. We note that a particle carry $l_I$ $a^I_\mu$-charge created a
$l_2\pi$ flux in $a^1_\mu$.  So a unit $a^2_\mu$-charge always create a
$G_s$-twist.  But what is the $G_s$-charge of the $l_I$ particle?  

To measure the $G_s$-charge, we need to find the pure $G_s$-twist.  Let us
assume that the pure $G_s$-twist corresponds to $\v l^v=(l^v_1,l^v_2,0,0)$
$a^I_\mu$-charge. Then $l^v_2=1$ so that the $\v l^v$ particle produce
$\pi$ $a^1_\mu$-flux.  
For a pure $G_s$-twist, we also have
\begin{align}
 \pi (\v l^v)^T K^{-1} \v l^v =0.
\end{align}
This allows us to obtain
\begin{align}
  (\v l^v)^T= (\frac{n_{1}}{2},1,0, 0 ). 
\end{align}
Note that some times, $\v l^v$ is not a allowed excitation. But we can always
use $\v l^v$ to probe the $G_s$ charge.  Let
\begin{align}
 \v q= 2K^{-1} \v l^v = \begin{pmatrix}
 1\\
 -n_1/2\\
 0\\
 -n_{12}/2\\
\end{pmatrix}.
\end{align}
Moving a  pure $G_s$-twist around the $l_I$ particle
will induce a phase
\begin{align}
 2\pi \v l^T K^{-1} \v l^v=\pi \v q^T \v l.
\end{align}
We find that the $G_s$-charge of the $l_I$ particle is
\begin{align}
\label{Gsn1n2n3}
 G_s\text{-charge}= \v q^T \v l \text{ mod } 2.
\end{align}

When $n_{12}=0$, those gauge
excitations have a trivial  mutual statistics with the  unit $a^2_\mu$-charge
(\ie the end of branch-cut).  This means that those  gauge excitations carry a
trivial $G_s$ quantum number.  When  $n_{12}=1$, the unit $a^4_\mu$-charge (the
gauge-flux excitation) has a $\pi/2$ mutual statistics with the  unit
$a^2_\mu$-charge (\ie the end of branch-cut).  This means that the  unit
$a^4_\mu$-charge carries a \emph{fractional} $G_s$ charge!  Such a
fractional-$G_s$-charge gauge excitation has a Bose/Fermi statistics if $n_2=0$
and a semion statistics if $n_2=1$.  We see that both $n_{12}$ and $n_2$ are
measurable.  $n_1$ is also measurable which describes the $G_s$ SPT phases.

To summarize, tables \ref{Z2tb1}--\ref{Z2tb8} list the  $G_s$-charges, the
$G_s$-twists, the $G_g$ gauge sectors, and the statistics of the 16 kinds of
quasiparticles/defects in the $Z_2$ gauge theory which contains a topological
term labeled by $n_1$, $n_{12}$, and $n_2$.  The  $G_s$-charge is a
$Z_2$-charge which is defined modular 2.  The $G_s$-twist = 0 means that there
is no branch-cut, and the $G_s$-twist = 1 means that there is a branch-cut with
the $G_s$ twist.  The statistics in tables \ref{Z2tb1}--\ref{Z2tb8} is defined
as statistics = $\th_l/\pi$.  Thus  statistics = 0 corresponds to Bose
statistics, statistics = 1 corresponds to Fermi statistics, and statistics =
$\pm 1/2$ correspond to semion statistics, etc.

The $G_g$ gauge excitations must have trivial mutual statistics with the $G_s$
charge and are described by $(l_I)=(0,0,l_3,l_4)$.  The
$G_g$-gauge sectors describe the four types
of $G_g$ gauge excitations:\\
the trivial excitation $(l_3,l_4)=(0,0) \to$ ``0'',\\
the $G_g$-charge excitation $(l_3,l_4)=(1,0) \to$ ``e'',\\
the $G_g$-vortex excitation $(l_3,l_4)=(0,1) \to$  ``m'', \\
the $G_g$-charge-vortex excitation $(l_3,l_4)=(1,1) \to$  ``em''.\\

We know that the above 8 classes of SET states are classified by
\begin{align}
&\ \ \ \
 \cH^3(Z_2\times Z_2,\R/\Z)
\nonumber\\
&=
 \cH^3(G_s=Z_2,\R/\Z)\oplus
 \cH^3(G_g=Z_2,\R/\Z)\oplus
\nonumber\\
& 
\ \ \ \ \ \ \ \ \ \ \ \
\ \ \ \ \ \ \ \ \ \ \ \
 \cH^2(G_s=Z_2,\Z_2)
\nonumber\\
&
=
\Z_2^3,
\end{align}
From the tables \ref{Z2tb1}--\ref{Z2tb8}, we see that
$\cH^3(G_g=Z_2,\R/\Z)=\Z_2$ (labeled by $n_2$) determine if the $G_g$ gauge
theory is a $Z_2$ gauge theory (for $n_2=0$) or a double-semion theory (for
$n_2=1$).  We also see that $ \cH^3(G_s=Z_2,\R/\Z)=\Z_2$ (labeled by $n_1$)
describes the $G_s$ SPT phases, and $\cH^2(G_s=Z_2,\Z_2)=\Z_2$ (labeled by
$n_{12}$) determine if the $G_g$ gauge-flux excitations can carry a 1/2 $G_s$
charge.

From the tables \ref{Z2tb1}--\ref{Z2tb8}, we see that some times, a 1/2 $G_s$
charge can and can only appear on a gauge-flux excitation with $l_4=1$. This
implies that the symmetry of the  gauge-flux excitations is described by a
non-trivial PSG $=Z_4$.  In all the 8 phases, the $G_g$ gauge-charge
excitations (the $a^3_\mu$-charges) are always bosonic and always carry integer
$G_s$ charge.  In other words, the symmetry of the gauge-charge excitations is
described by a trivial PSG $=G_s\times G_g=Z_2\times Z_2$.

Next, we consider the $G=Z_4$ case.  We will show that, in this case, the
symmetry of the gauge-charge excitations is described by a non-trivial PSG
$=Z_4$ (\ie carries a fractional $G_s$-charge).  A $G=Z_4$ gauge theory can be
described by $U^2(1)$ mutual Chern-Simons theory:
\begin{align}
 \cL=\frac{1}{4\pi} K_{0,IJ} a^I_\mu\prt_\nu a^J_\la + ...
\end{align}
with
\begin{align}
 K_0=4\bpm
0&1\\
1&0\\
\epm
\end{align}
A unit $G_g$ gauge-charge corresponds to the unit charge of $a^1_\mu$ gauge
field and a $G_g$ gauge-flux excitation corresponds to two-unit charge of
$a^2_\mu$ gauge field.  Note that a unit $G_g$ gauge-charge carries 1/2 $G_s$
charge!  In other words, the symmetry of the gauge-charge  excitations is
described by a non-trivial PSG $=Z_4$.  Two-unit charge of $a^1_\mu$ gauge
field carries no $G_g$ gauge-charge, but a unit of $G_s$ charge.

The 4 types of quantized topological terms are given by
\begin{align}
 W_\text{top}=
\frac{m_1}{2\pi}  a^1_\mu\prt_\nu a^1_\la
\end{align}
$m_1=0,1,2,3$.
The total Lagrangian has a form
\begin{align}
 \cL+W_\text{top}=\frac{1}{4\pi} K_{IJ} a^I_\mu\prt_\nu a^J_\la + ...
\end{align}
with
\begin{align}
 K=\bpm
2m_1&4\\
4&0\\
\epm ,\ \ \
 K^{-1}=\frac 18 \bpm
0&2\\
2&-m_1\\
\epm .
\end{align}
Since moving the $G_s$ charge (two units of $a^1_\mu$-charge) around a
unit-$a^2_\mu$-charge induced a phase $\pi$, a unit $a^2_\mu$-charge correspond
to the end of branch-cut in the original theory along which we have a $G_s$
symmetry twist.  However, fusing two unit-$a^2_\mu$-charge give a non-trivial
$G_g$ gauge excitation -- a unit of $G_g$ gauge flux (described by two-unit
charge of $a^2_\mu$ gauge field).  Therefore a unit $a^2_\mu$-charge does not
correspond to a pure $G_s$ twist.  It is a bound state of $G_s$ twist, $G_g$
gauge excitation, and $G_s$ charge.  

To calculate the  $G_s$ charge for a  generic quasiparticle with $l_I$
$a^I_\mu$-charge, first we assume that that  the $G_s$ charge has the following
form
\begin{align}
 G_s\text{-charge}=\v l^T \v q.
\end{align}
The vector $\v q$ must satisfy $(2,0) \v q =\pm 1$ so that two units of
$a^1_\mu$-charge carry a $G_s$ charge 1.  To obtain another condition on $\v
q$, we note that the trivial quasiparticles are given by $\v
l=(K_{11},K_{12})=(2m_1, 4)$ and $\v l=(K_{21},K_{22})=(4,0)$.  So we require
that $(2m_1, 4) \v q = 0$ or $2$.
We find that $\v q$ has four choices
\begin{align}
 \v q^T &=(1/2,-m_1/4), &  \v q^T &=(-1/2,m_1/4), 
\\
 \v q^T &=(1/2,(2-m_1)/4), &  \v q^T &=(-1/2,(2+m_1)/4).
\nonumber 
\end{align}

We may choose $\v q^T=(1/2,-m_1/4)$ and obtain tables \ref{Z4tb1}-\ref{Z4tb4},
which list the  $G_s$-charges, the $G_s$-twists, the $G_g$ gauge sectors, and
the statistics of the 16 kinds of quasiparticles/defects in the $Z_2$ gauge
theory with $Z_2$ symmetry which contain a topological term labeled by $m_1$
and a mixing of the gauge $G_g$ and symmetry $G_s$ described by $G=Z_4$.  Other
choices of  $\v q$ sometimes regenerate the above four states and sometimes
generate new states.

From tables \ref{Z2tb1}--\ref{Z4tb4}, we see the patterns of $G_s$-charges,
$G_s$ twists, and statistics are all different, except the
$(n_1n_{12}n_2)=(010)$ state and the $m_1=0$ state: the two states are related
by an exchange $e \leftrightarrow m$.  Thus the construction produces 11
different  $Z_2$ gauge theories with $Z_2$ symmetry.  

Let us examine the quasiparticles without the
$G_s$-twist.  We see 6 states contain quasiparticles with bosonic and fermionic
statistics.  Those 6 states are described by standard $G_g=Z_2$ gauge theory.
However, the $G_s=Z_2$ symmetry is realized differently.  Some states contain
quasiparticles with fractional $G_s=Z_2$ charge while others without fractional
$G_s=Z_2$ charge.  In some states, the fermionic quasiparticles carry
fractional $G_s=Z_2$ charge while in other states,  the fermionic
quasiparticles carry integer $G_s=Z_2$ charge.

The other 6 states contain  quasiparticles with semion statistics.  Those
states are twisted $Z_2$ gauge theory which is also known as double-semion
theory.\cite{FNS0428,LW0510} Again some of those states have fractional
$G_s=Z_2$ charge while others without fractional $G_s=Z_2$ charge.  Some times,
the semions only carry integer  $G_s=Z_2$ charges, or only  fractional
$G_s=Z_2$ charges, or both integer and fractional $G_s=Z_2$ charges.
Those results agree with those obtained in \Ref{LV1334,HW1351}.

\subsection{Comparison with group cohomology construction}

In \Ref{MR1235}, SET phases are constructed using group cohomology,
generalizing the Toric code to include global symmetry. The physical
excitations in phases with the group extension given by $G=G_s \times G_g
=\Z_2\times \Z_2$ were also explored there, and it is of interest to compare
with the results above using $K$-matrix.

The group cohomology $\cH^3(\Z_2\times\Z_2, \R/\Z)=
\Z_2\times\Z_2\times\Z_2$.  The generators of each of the $\Z_2$ in the
cohomology group is given by
\begin{eqnarray}
\omega_{11}(x,y,z) &=& \exp \left(\frac{\pi i}{2}x_1 (y_1+ z_1 - \overline{y_1+z_1})\right) \\
\omega_{22}(x,y,z) &=&   \exp \left(\frac{\pi i}{2}x_2 (y_2+ z_2 - \overline{y_2+z_2})\right)\\
\omega_{12}(x,y,z) &=&\exp \left(\frac{\pi i}{2}x_1 (y_2+ z_2 - \overline{y_2+z_2})\right)
\end{eqnarray}
where $x,y,z \in \Z_2 \times\Z_2$, and $x= (x_1,x_2)$ where $x_{1,2}=\{0,1\}$,
and similarly for $y$ and $z$. Also $\overline{a+b} = a+b \,\,\bmod\,2$.
Note that
\begin{align}
&\ \ \ \
 \cH^3(\Z_2\times\Z_2, \R/\Z)
\nonumber\\ &=
 \cH^3[\Z_2, \R/\Z)\oplus
 \cH^2[\Z_2, \cH^1(\Z_2, \R/\Z)]\oplus
\nonumber\\ &\ \ \ \
 \cH^1[\Z_2, \cH^2(\Z_2, \R/\Z)]\oplus
 \cH^3(\Z_2, \R/\Z)]
\nonumber\\ &=
\Z_2\oplus \Z_2\oplus \Z_1\oplus \Z_2
\nonumber\\ &=
\Z_2\times \Z_2\times \Z_1\times \Z_2
\end{align}

A phase is then characterized by three-cocycles of the form
\be
\Omega(x,y,z) = \omega_{11}^{n_1}(x,y,z) \omega_{22}^{n_2}(x,y,z)\omega_{12}^{n_{12}}(x,y,z),
\ee
where $n_{1,12,2}=\{0,1\}$, and they can be precisely identified with the
$n_1,\ n_{12},\ n_2$ in \eqn{csz2z2}. This can be easily checked by computing
the modular $S$-matrix from the group cycles, and comparing with the matrix of
mutual statistics obtained from the $K$-matrix.
More explicitly, using the methods detailed in \Ref{DW9093,DVV8985,HWW1295},
the modular $S$-matrix evaluated on the cocycle $\Omega(x,y,z)$ of the
$\Z_2\times\Z_2$ lattice gauge theory is given by
\begin{eqnarray}\label{modS}
&&\ \ \ S_{(g,\alpha)(h,\beta)}(n_1,n_{12},n_2)  \nonumber\\
&&= \frac{1}{4} \exp\bigg(-\pi i ([\sum_i^2 \alpha_i h_i + \beta_i g_i] \\
&&\ \ \ \ \ \ \ \ \ \ \ \ \ \
+ n_1 g_1 h_1 + n_2 g_2 h_2 + \frac{n_{12}}{2}(g_1h_2+h_1g_2))\bigg)
\nonumber
\end{eqnarray}
where $g,h,\alpha,\beta $ are all two component vectors
whose components each taking values $\in \{0,1\}$. Here
$g,h \in \Z_2\times\Z_2$ are the flux excitations,
and $\alpha,\beta$ denote irreducible representations of
$\Z_2\times\Z_2$, which correspond to charge excitations.
The phase factor appearing in the modular matrix is related to
the mutual statistics obtained in \eqn{mutual}.
It is clear that the phase factor indeed takes the form of
\eqn{mutual} if we interpret $(\alpha_1, g_1,\alpha_2,g_2)$
and $(\beta_1, h_1,\beta_2,h_2)$
as our charge vectors $l,l'$ respectively:\cite{Wrig,KW9327}
\begin{eqnarray}
S_{l,l'}(n_1,n_{12},n_2)  = \frac{1}{4} \exp\bigg(-2\pi i l^T K^{-1} l'\bigg).
\end{eqnarray}
 We can thus immediately read
off the inverse of the $K$-matrix from \eqn{modS} to be
\be
K^{-1} = \frac14\left(\begin{tabular}{cccc}
0&2&0&0\\
2&2$n_1$&0&$n_{12}$\\
0&0&0&2\\
0&$n_{12}$&2&2$n_2$
\end{tabular}
\right),
\ee
where up to a convention for the sign of $n_1,\ n_{12},\ n_3$ is precisely \eqn{csz2z2}.

In \Ref{MR1235} the $G_s$ charges of both flux and charge excitations of the
gauge group $G_g$ are computed, by explicitly constructing the $G_s$ symmetry
transformation operator and the (pair) creation operators (\ie ribbon
operators) of the excitations. In the language of the $K$-matrix construction,
the gauge-charge and flux excitations correspond to charges of $a_3$ and $a_4$
respectively.  \ie violation of vanishing flux in a plaquette corresponds to
$a_4$ charges, and the $a_3$ charges correspond to the product of gauge
variables along the ribbon connecting the pair of excitations at the end points
of the ribbon.  $G_s$ charge fluctuations are also possible in the cocycle
model, but it does not contain $G_s$-flux excitation by construction there. An
$a_2$ charge would correspond to a field configuration in \Ref{MR1235} which
does not return to its original value after traversing a loop.  Therefore we
can compare the $G_s$ charges of excitations with those in \Ref{MR1235} when
$l_2=0$.

Let us elaborate further on the conversion of gauge charges between the two
descriptions. In \Ref{MR1235} excited states with a pair of quasi-particle
excitations are specified by $|h,h_g,\tilde{g},u_A\rangle$, where $h,h_g\in G_g$,
$\tilde{g}, u_A\in G_s$, and $u_A$ corresponds to the field configuration at
one of the two quasi-particle sites $A,B$ connected by the ribbon operator. It
satisfies the constraint $u_A u_B^{-1}= \tilde{g}$.  Flux excitations are given
by $h$, whereas charge fluctuations are given by $h_g$, and $G_s$ charges are
given by a mixture of $\tilde{g}, u_A$. The charge fluctuations are however
expressed in a different basis compared to the $K$-matrix description.  To
convert to the $K$ matrix description, we again have to do the following
transformation (suppose we focus on the quasiparticle located at the end $B$,
and fixing $u_A$ at the other end) \be |h,\alpha_g,\beta_s,u_A\rangle =
\frac{1}{|G_g\times G_s|}\sum_{h_g,\tilde{g}}\rho_{\alpha_g}(h_g)
\rho_{\beta_s} (\tilde{g})|h,h_g,\tilde{g},u_A\rangle , \ee where
$\rho_{\alpha_g}(g)$ corresponds to characters of representations of
$G_g=\Z_2$, and $\rho_{\beta_s}(\tilde{g})$ that of $G_s=\Z_2$.\footnote{In the
  case of non-vanishing $n_{12}$, the charges of $G_s$ and $G_g$ mix, and the
basis becomes a linear combinations of the above.} One can check that in terms
of the diagonalized basis vectors of the $G_s$ transformation as specified in
Table II in \Ref{MR1235}, the $G_s$ charge match up with the result obtained in
the $K$-matrix formulation given above.

The most important observation is that it is found in \Ref{MR1235} (see table
II there) that only in the case where $n_{12}$ and $l_4$ (\ie flux charge $h=1$
there) are \emph{both} non-vanishing that charge fractionalization occurs. In
fact the $G_s$ transformation $U$ for the flux charge squares to $-1$, which is
indeed the statement that the $G_s$ charge is halved. This is in perfect
agreement with the results in the previous section (see \eqn{Gsn1n2n3} or
tables \ref{Z2tb1}--\ref{Z2tb8}).

We note also that since the modular $S$-matrix descending from the 3-cocycles
agree with that of the $K$-matrix, the braiding statistics in \Ref{MR1235} have
to agree with that obtained using the $K$-matrix when we turn off $l_2$
accordingly.

\section{Summary}

In this paper, we studied the quantized topological terms in a weak-coupling
gauge theory with gauge group $G_g$ and a global symmetry $G_s$ in
$d$-dimensional space-time. We showed that the  quantized topological terms are
classified by a pair $(G,\nu_d)$, where $G$ is an extension of $G_s$ by $G_g$
and $\nu_d$ is an element in group cohomology $\cH^d(G,\R/\Z)$.  When $d=3$
and/or when $G_g$ is finite, the  weak-coupling gauge theories with  quantized
topological terms describe gapped SET phases.  Thus those SET phases are
classified by $\cH^d(G,\R/\Z)$, where $G/G_g=G_s$.  This result generalized the
PSG description of the SET phases.\cite{W0213,W0303a,KLW0834,KW0906}. It also
generalized the recent results in \Ref{EH1293,MR1235}.  We also apply our
theory to a simple case $G_s=G_g=Z_2$, to understand the physical meanings of
the  $\cH^d(G,\R/\Z)$ classification.  Roughly, for the trivial extension
$G=G_s\times G_g$, $\cH^d(G_g\times G_s,\R/\Z)$ describes different ways in
which the quantum number of $G_s$ becomes fractionalized on gauge-flux
excitations.  While the non-trivial extensions $G$ describe different ways in
which the quantum number of $G_s$ become fractionalized on gauge-charge
excitations.

We like to thank Y.-M. Lu and Ashvin Vishwanath for discussions.  This research
is supported by NSF Grant No. DMR-1005541, NSFC 11074140, and NSFC 11274192.
Research at Perimeter Institute is supported by the Government of Canada
through Industry Canada and by the Province of Ontario through the Ministry of
Research. LYH is supported by the Croucher Fellowship.

\appendix

\section{ Calculate  $H^*(X,\R/\Z)$ from $H^*(X,\Z)$ } \label{HBGRZ}

We can use the K\"unneth formula (see \Ref{Spa66} page 247)
\begin{align}
\label{kunn}
&\ \ \ \ H^d(X\times X',M\otimes_R M')
\nonumber\\
&\simeq \Big[\oplus_{p=0}^d H^p(X,M)\otimes_R H^{d-p}(X',M')\Big]\oplus
\nonumber\\
&\ \ \ \ \ \
\Big[\oplus_{p=0}^{d+1}
\text{Tor}_1^R(H^p(X,M),H^{d-p+1}(X',M'))\Big]  .
\end{align}
to calculate $H^*(X,M)$ from $H^*(X,Z)$.  Here $R$ is a principle ideal domain
and $M,M'$ are $R$-modules such that $\text{Tor}_1^R(M,M')=0$.  Note that $\Z$
and $\R$ are principal ideal domains, while $\R/\Z$ is not.  A $R$-module is
like a vector space over $R$ (\ie we can ``multiply'' a vector by an element of
$R$.) For more details on principal ideal domain and $R$-module, see the
corresponding Wiki articles.

The tensor-product operation $\otimes_R$ and  the
torsion-product operation $\text{Tor}_1^R$ have the following properties:
\begin{align}
\label{tnprd}
& A \otimes_\Z B \simeq B \otimes_\Z A ,
\nonumber\\
& \Z \otimes_\Z M \simeq M \otimes_\Z \Z =M ,
\nonumber\\
& \Z_n \otimes_\Z M \simeq M \otimes_\Z \Z_n = M/nM ,
\nonumber\\
& \Z_m \otimes_\Z \Z_n  =\Z_{(m,n)} ,
\nonumber\\
&  (A\oplus B)\otimes_R M = (A \otimes_R M)\oplus (B \otimes_R M)   ,
\nonumber\\
& M \otimes_R (A\oplus B) = (M \otimes_R A)\oplus (M \otimes_R B)   ;
\end{align}
and
\begin{align}
\label{trprd}
& \text{Tor}_1^R(A,B) \simeq \text{Tor}_1^R(B,A)  ,
\nonumber\\
& \text{Tor}_1^\Z(\Z, M) = \text{Tor}_1^\Z(M, \Z) = 0,
\nonumber\\
& \text{Tor}_1^\Z(\Z_n, M) = \{m\in M| nm=0\},
\nonumber\\
& \text{Tor}_1^\Z(\Z_m, \Z_n) = \Z_{(m,n)} ,
\nonumber\\
& \text{Tor}_1^R(A\oplus B,M) = \text{Tor}_1^R(A, M)\oplus\text{Tor}_1^R(B, M),
\nonumber\\
& \text{Tor}_1^R(M,A\oplus B) = \text{Tor}_1^R(M,A)\oplus\text{Tor}_1^R(M,B)
,
\end{align}
where $(m,n)$ is the greatest common divisor of $m$ and $n$.  These expressions
allow us to compute the tensor-product $\otimes_R$ and  the torsion-product
$\text{Tor}_1^R$.

If we choose $R=M=\Z$, then the condition
$\text{Tor}_1^R(M,M')=\text{Tor}_1^{\Z}(\Z,M')=0$ is always satisfied. So we
have
\begin{align}
\label{kunnZ}
&\ \ \ \ H^d(X\times X',M')
\nonumber\\
&\simeq \Big[\oplus_{p=0}^d H^p(X,\Z)\otimes_{\Z} H^{d-p}(X',M')\Big]\oplus
\nonumber\\
&\ \ \ \ \ \
\Big[\oplus_{p=0}^{d+1}
\text{Tor}_1^{\Z}(H^p(X,{\Z}),H^{d-p+1}(X',M'))\Big]  .
\end{align}
Now we can further choose $X'$ to be the space of one point, and use
\begin{align}
H^{d}(X',M'))=
\begin{cases}
M', & \text{ if } d=0,\\
0, & \text{ if } d>0,
\end{cases}
\end{align}
to reduce \eqn{kunnZ} to
\begin{align}
\label{ucf}
&\ \ \ \ H^d(X,M)
\\
&\simeq  H^d(X,\Z)\otimes_{\Z} M \oplus
\text{Tor}_1^{\Z}(H^{d+1}(X,{\Z}),M)  ,
\nonumber
\end{align}
where $M'$ is renamed as $M$.  The above is a form of the universal coefficient
theorem which can be used to calculate $H^*(BG,M)$ from $H^*(BG,\Z)$ and the
module $M$.

Now, let us choose $M=\R/\Z$ and compute
$H^d (BG,\R/\Z)$ from $H^d (BG,\Z)$.  Note that $H^d (BG,\Z)$ has a form $H^d
(BG,\Z)=\Z\oplus ... \oplus \Z \oplus Z_{n_1} \oplus Z_{n_2}\oplus ...  $.  A
$\Z$ in $H^d (BG,\Z)$ will produce a $\R/\Z$ in $H^d (BG,\R/\Z)$ since $\Z
\otimes_{\Z} \R/\Z=\R/\Z$.  A $\Z_n$ in $H^{d+1} (BG,\Z)$ will produce a $\Z_n$
in $H^d (BG,\R/\Z)$ since $\text{Tor}_1^{\Z}(\Z_n, \R/\Z)=\Z_n$.  So we see
that $H^d (BG,\R/\Z)$ has a form $H^d (BG,\R/\Z)=\R/\Z\oplus ... \oplus \R/\Z
\oplus Z_{n_1} \oplus Z_{n_2}\oplus ...  $ and
\begin{align}
\label{DisH}
\text{Dis}[ H^d (X,\R/\Z) ] \simeq \text{Tor}[H^{d+1}(X,\Z)] .
\end{align}
where $\text{Dis}[ H^d (X,\R/\Z) ]$ is the discrete part of $H^d (X,\R/\Z)$.

If we choose $M=\R$, we find that
\begin{align}
\label{ucfR}
 H^d(X,\R)
\simeq  H^d(X,\Z)\otimes_{\Z} \R.
\end{align}
So $ H^d(X,\R)$ has the form $\R\oplus ... \oplus \R$ and each $\Z$ in
$H^d(X,\Z)$ gives rise to a $\R$ in $ H^d(X,\R)$.
Since $ H^d(BG,\R)=0$ for $d =$ odd,
we have
\begin{align}
\label{HTorH}
  H^d(BG,\Z)=\text{Tor}[ H^d(BG,\Z)],\ \ \text{ for } d= \text{ odd}.
\end{align}

Using the K\"unneth formula \eqn{kunnZ}
we can also rewrite $H^{d}(G_s\times G_g,\R/\Z)$
as
\begin{align}
\label{GsGg}
&\ \ \ \ \cH^{d}(G_s\times G_g,\R/\Z)
\nonumber\\
& = \cH^{d+1}(G_s\times G_g,\Z)
\nonumber\\
& =
\Big[ \oplus_{p=0}^{d+1} \cH^p(G_s,\Z)\otimes_{\Z} \cH^{d+1-p}(G,\Z)
\Big]\oplus
\nonumber\\
&\ \ \ \
\Big[\oplus_{p=0}^{d+2}
\text{Tor}_1^{\Z}[\cH^p(G_s,{\Z}),\cH^{d-p+2}(G,\Z)]\Big]
\nonumber\\
& = \cH^d(G_s,\R/\Z)\oplus  \cH^{d}(G_g,\R/\Z)\oplus
\nonumber\\
&\ \ \ \
\Big[ \oplus_{p=1}^{d-1} \cH^{d-p}(G_s,\Z)\otimes_{\Z} \cH^{p}(G_g,\R/\Z)
\Big]\oplus
\nonumber\\
&\ \ \ \
\Big[\oplus_{p=1}^{d-1}
\text{Tor}_1^{\Z}[\cH^{d-p+1}(G_s,\Z),\cH^{p}(G_g,\R/\Z)]\Big]
\nonumber\\
& = \cH^d(G_s,\R/\Z)\oplus  \cH^{d}(G_g,\R/\Z)\oplus
\nonumber\\
&\ \ \ \
\Big[ \oplus_{p=1}^{d-1} \cH^{d-p}[G_s,\cH^{p}(G_g,\R/\Z)]
\Big]
\nonumber\\
& =
 \oplus_{p=0}^{d} \cH^{d-p}[G_s,\cH^{p}(G_g,\R/\Z)]
,
\end{align}
where we have used $\cH^n(G,\R/\Z)=\cH^{n+1}(G,\Z)$
for $n>0$, and $\cH^{1}(G,\Z)=0$ for compact or finite group $G$.
We also used the
universal coefficient
theorem \eq{ucf}
\begin{align}
&\ \ \ \
  \cH^{d-p}[G_s,\cH^{p}(G_g,\R/\Z)]
\nonumber\\
&= \cH^{d-p}(G_s,\Z)\otimes_{\Z} \cH^{p}(G_g,\R/\Z)
\oplus
\nonumber\\
&\ \ \ \
\text{Tor}_1^{\Z}[\cH^{d-p+1}(G_s,\Z),\cH^{p}(G_g,\R/\Z)]
\end{align}

\section{A labeling scheme of SET states
described by weak-coupling gauge theory}
\label{lab}

The Lyndon-Hochschild-Serre spectral sequence $ \cH^x[G_s,\cH^y(G_g,\R/\Z)]
\Rightarrow \cH^{x+y}(G,\R/\Z)$ may help us to calculate the  group cohomology
$\cH^d(G,\R/\Z)$ in terms of  $\cH^x[G_s,\cH^y(G_g,\R/\Z)]$.  We find that
$\cH^d(G,\R/\Z)$ contains a chain of subgroups
\begin{align}
\{0\}=H^{d+1}
\subset H^d
\subset ...
\subset H^1
\subset H^0
=
 \cH^d(G,\R/\Z)
\end{align}
such that $H^k/H^{k+1}$ is a subgroup of a factor
group of $\cH^k[G_s,\cH^{d-k}(G_g,\R/\Z)]$:
\begin{align}
 H^k/H^{k+1} \subset \cH^k[G_s,\cH^{d-k}(G_g,\R/\Z)]/\cH^k,\ \
k=0,...,d,
\end{align}
where $\cH^k$ is a subgroup of $\cH^k[G_s,\cH^{d-k}(G_g,\R/\Z)]$. Note that
$G_s$ has a non-trivial action on $\cH^{d-k}(G_g,\R/\Z)$ as determined by the
structure $G_s=G/G_g$.  We also have
\begin{align}
 H^0/H^{1} &\subset \cH^0[G_s,\cH^{d}(G_g,\R/\Z)],
\nonumber\\
 H^d/H^{d+1}&=H^d = \cH^d(G_s,\R/\Z)/\cH^d.
\end{align}
In other words, the elements in $\cH^d(G,\R/\Z)$ can be one-to-one labeled by
$(x_0,x_1,...,x_d)$ with
\begin{align}
 x_k\in H^k/H^{k+1} \subset \cH^k[G_s,\cH^{d-k}(G_g,\R/\Z)]/\cH^k.
\end{align}
If we want to use $(y_0,y_1,...,y_d)$ with
\begin{align}
 y_k\in \cH^k[G_s,\cH^{d-k}(G_g,\R/\Z)]
\end{align}
to  label the  elements in $\cH^d(G,\R/\Z)$, then such a labeling may not be
one-to-one and it may happen that only some of $(y_0,y_1,...,y_d)$ correspond
to  the  elements in $\cH^d(G,\R/\Z)$.  But for every element in
$\cH^d(G,\R/\Z)$, we can find a $(y_0,y_1,...,y_d)$ that corresponds to it.

\bibliography{../../../bib/wencross,../../../bib/all,../../../bib/publst,./tmp}

\end{document}